\documentclass[journal=jacsat,manuscript=article]{achemso}
\setkeys{acs}{usetitle = true}

\usepackage[version=3]{mhchem} 
\usepackage{graphicx} 
\usepackage{dcolumn}  
\usepackage{amssymb,latexsym}
\usepackage{bm}
\usepackage{amsbsy} 
\usepackage{amsmath}
\usepackage{color}
\usepackage{subfigure}
\usepackage{cleveref}
\usepackage{ dsfont }
\usepackage{physics}
\usepackage{nicefrac}

\author{Matthew S. Ryley}
\affiliation{School of Chemistry, University of Nottingham, University Park, Nottingham, NG7 2RD, UK}
\author{Michael Withnall}
\affiliation{School of Chemistry, University of Nottingham, University Park, Nottingham, NG7 2RD, UK}
\author{Tom J. P. Irons}
\affiliation{School of Chemistry, University of Nottingham, University Park, Nottingham, NG7 2RD, UK}
\author{Trygve Helgaker}
\affiliation{Hylleraas Centre for Quantum Molecular Sciences, Department of Chemistry, University of Oslo, P.O. Box 1033 Blindern, N-0315 Oslo, Norway}
\author{Andrew M. Teale}
\email{andrew.teale@nottingham.ac.uk}
\phone{+44 (0)115 846 8482}
\affiliation{School of Chemistry, University of Nottingham, University Park, Nottingham, NG7 2RD, UK}
\alsoaffiliation{Hylleraas Centre for Quantum Molecular Sciences, Department of Chemistry, University of Oslo, P.O. Box 1033 Blindern, N-0315 Oslo, Norway}

\title{Robust All-Electron Optimization in Orbital-Free Density-Functional Theory Using the Trust-Region Image Method}

\begin{document}
\begin{abstract}
We present a Gaussian-basis implementation of orbital-free density-functional theory (OF-DFT) in which the trust-region image method (TRIM) is used for optimization. This second-order optimization scheme has been constructed to provide benchmark all-electron results with very tight convergence of the particle-number constraint, associated chemical potential and electron density. It is demonstrated that, by preserving the saddle-point nature of the optimization and simultaneously optimizing the density and chemical potential, an order of magnitude reduction in the number of iterations required for convergence is obtained. The approach is compared and contrasted with a new implementation of the nested optimization scheme put forward by Chan, Cohen and Handy. Our implementation allows for semi-local kinetic-energy (and exchange--correlation) functionals to be handled self-consistently in all-electron calculations. The all-electron Gaussian-basis setting for these calculations will enable direct comparison with a wide range of standard high-accuracy quantum-chemical methods as well as with Kohn--Sham density-functional theory. We expect that the present implementation will provide a useful tool for analysing the performance of approximate kinetic-energy functionals in finite systems.       
\end{abstract}

\maketitle

\section{Introduction}\label{sec:intro}
The Kohn--Sham formulation of density-functional theory~\cite{Kohn1965} (DFT) has enjoyed enormous success and widespread use since its introduction in 1965. In this approach, an auxiliary, noninteracting system of electrons is introduced with the same electron density as that of the physical system. This auxiliary system is associated with a single-determinant wave function and set of orbitals. A major contribution to the accuracy, and hence success, of the theory is that this approach facilitates the description of a major component of the kinetic energy $T$ by the noninteracting kinetic energy $T_\mathrm{s}$. Whilst $T$ and $T_\mathrm{s}$ are formally functionals of the electron density $\rho$, their explicit forms are unknown. In Kohn--Sham DFT (KS-DFT), $T_\mathrm{s}$ is evaluated exactly as an explicit functional of the occupied molecular orbitals or, equivalently, the one-particle reduced density matrix $\rho(\mathbf{r},\mathbf{r}^\prime)$ -- that is,  $T_\mathrm{s}(\rho(\mathbf{r},\mathbf{r}^\prime))$.       

The use of $T_\mathrm{s}(\rho(\mathbf{r},\mathbf{r}^\prime))$ in the Kohn--Sham scheme neatly sidesteps the need to develop approximations for $T_\mathrm{s}(\rho)$; the only term which must be approximated is the exchange--correlation energy $E_\mathrm{xc}(\rho)$, to which progressively more accurate approximations with a favourable balance between computational cost and accuracy have been developed over the previous five decades~\cite{Burke2012,Cohen2012,Becke2014,Jones2015}. However, the introduction of a noninteracting wavefunction is not without cost. The resulting Kohn--Sham equations are a set of $N$ coupled equations for the required orbitals, where $N$ is the number of electrons in the system. Impressive progress has been made in recent years towards $\mathcal{O}(N)$ implementation of Kohn--Sham methods~\cite{Johnson1996,Goedecker1999,Scuseria1999,Wu2002,Hine2009,ochsenfeld2007linear,Neese2011,Rubensson2011,Bowler2012}, which introduce numerous approximations to accelerate the evaluation of the molecular integrals, avoid costly diagonalization steps in the solution of the Kohn--Sham equations and harness large-scale parallel computers. As a result, the size of system that can be modelled has increased dramatically in recent years. 

Whilst these developments have helped enable the application of KS-DFT methods to ever larger systems, a further significant increase in the size of system and speed of calculations is possible if the use of orbitals can be eliminated entirely. Orbital-free density-functional theories (OF-DFTs) that provide sufficient accuracy for chemical systems have been a goal that significantly predates Hohenberg and Kohn's landmark 1964 paper~\cite{Hohenberg1964}, originating from the early models of Thomas and Fermi in 1927~\cite{Thomas1927,Fermi1927} and Dirac in 1930~\cite{Diracfunctional}. Orbital-free approaches can be readily formulated in a linear-scaling manner and, since only a single Euler--Lagrange equation defining the electronic density must be solved, the pre-factor of such an approach compared with KS-DFT is significantly reduced. 

Indeed, alongside developments in KS-DFT, programs such as PROFESS~\cite{Chen2015}, ATLAS~\cite{Mi2016}, DFTPy~\cite{DFTpy} and CONUNDrum~\cite{Golub2020} have been developed that apply OF-DFT to give a quantum-mechanical treatment of periodic systems with millions of atoms at modest computational cost. The potential efficiency of the approach is therefore clear, but a major outstanding challenge for OF-DFT is the development of accurate $T_\mathrm{s}(\rho)$.

However, in order to develop stable, accurate, transferable and practical noninteracting kinetic-energy functionals for use in OF-DFT, it is also necessary to be able to accurately and reliably determine self-consistent solutions to the Euler--Lagrange equation using these functionals. This is needed to establish not only a reasonable behaviour of the energy functional given accurate electron densities, but also that it has well-behaved functional derivatives and that optimization leads to stationary points that are physically meaningful. The most established OF-DFT codes make use of local pseudo-potentials and the extent to which these influence conclusions about the stability and accuracy of the underlying kinetic-energy functionals has been the subject of some discussion~\cite{Karasiev2012,Xia2015,TrickeyCommXia2015,ReplyXia2015}. 

Over the years, several attempts have been made to develop self-consistent all-electron OF-DFT codes. We note the work of Lopez-Acevedo and co-workers~\cite{lehtomaki2014orbital,leal2015optimizing}, where functionals comprising linear combinations of the Thomas--Fermi (TF)~\cite{Thomas1927,Fermi1927} and von-Weizs\"acker (vW)~\cite{VWFUNC} kinetic energies were considered. In the context of finite Gaussian-basis expansion methods, Chan, Cohen and Handy (CCH) presented an all-electron treatment based on direct optimization~\cite{Chan2001}. Although their approach is flexible enough to be applied with many energy functionals, they presented results only for $\gamma$TF$\lambda$vW type functionals. In general, the solution of the Euler--Lagrange equation has been reported to be challenging in the all-electron context without pseudo potentials -- see, for example, Refs.~\citenum{Chan2001,Karasiev2012}.

In this work, we aim to develop a flexible framework for the all-electron self-consistent solution of the Euler--Lagrange equation in OF-DFT with a range of energy functionals. We first give a brief overview of OF-DFT and the central optimization problem to be addressed. A new approach based on the trust-region image method (TRIM)~\cite{helgaker1991transition} is outlined, which exploits second-order information and offers quadratic convergence to the required self-consistent solutions. We then present some illustrative results, comparing the performance of this approach with the schemes of Lopez-Acevedo and co-workers~\cite{lehtomaki2014orbital} and CCH~\cite{Chan2001}. The importance of full self-consistency is highlighted when attempting to assess the quality of approximations to $T_\mathrm{s}(\rho)$. We conclude by discussing directions for future work based on the results of this work.

\section{Theory}\label{sec:theo}
\subsection{Four-way correspondence of OF-DFT}
\label{sec:4way}
In OF-DFT, the optimization of the ground-state energy is carried out using the electron density directly. To carry out this optimization, it is necessary to allow for variations with respect to the parameters defining the electron density subject to the constraint that the optimizing density is everywhere positive, has a finite kinetic energy and contains the correct number of electrons $N$. The appropriate context for describing the energy functional to be optimized in OF-DFT is grand-canonical ensemble (GCE) DFT, which allows for variation of the number of particles in the system. We do not give a comprehensive introduction to GCE-DFT here, but rather outline the key functionals of relevance to OF-DFT, the relationships between them, and their relevance for the practical optimization scheme to be introduced in this work. The fundamentals of GCE-DFT have long been established in DFT, though most calculations are carried out in the canonical ensemble. For introductions to GCE-DFT we refer the reader to Refs.~\citenum{Mermin1965,DFTYP,Eschrig1996}. Here we present the theory in terms of Lieb's convex-conjugate formulation of DFT~\cite{Lieb1983}, see for example Refs.~\citenum{Eschrig1996,Kutzelnigg2006,HJO} for introductions to this formulation of DFT. This formalism of GCE-DFT follows from the application of convex analysis, in particular for saddle-functions; a comprehensive exposition of the mathematical basis for this approach can be found in 
Refs.~\citenum{rockafellar1970convex,Rockafellar1968,rockafellar1971,HiriartUrruty1996,BertsekasNedic2003,BarbuPrecupano2012}.

The ground-state energy in the presence of an external potential $v$ and chemical potential $\mu \in \mathbb R$ is given by 
\begin{align}
E(v,\mu) &= \inf_{w_N,\psi_N} \sum_N w_N  \bigl\langle\psi_N \left\vert\, T_N + W_N + \textstyle \sum\nolimits_i {v_i}  - {\mu} N  \,\right\vert \psi_N \bigr\rangle  \label{eq:Evmu}
\end{align}
where $w_N \geq 0$, $\sum_N w_N = 1$, and $N$ is the particle number; $T_N$ and $W_N$ are the $N$--particle kinetic-energy and two-electron operators respectively and $\psi_N$ is a normalized $N$--particle wavefunction. It can be shown that $E(v,\mu)$ is separately concave and upper semi-continuous in the two variables $v$ and $\mu$. By the general theory of convex and concave functions, it follows that $E(v,\mu)$ is related to the concave--convex saddle function $\mathcal E(v,N)$, representing the energy of an $N$-electron system in the external potential $v$ as
\begin{align}
    E(v,\mu) &= \inf\nolimits_{N} \bigl(  \mathcal E(v,N) - \mu N\bigr), \label{eq:LFmuN}\\
    \mathcal E(v,N) &= \sup\nolimits_{\mu} \bigl( E(v,\mu) + \mu N\bigr) \label{eq:LFNmu}
\end{align}
and to the convex--concave saddle function $\mathcal H(\rho,\mu)$, representing the energy of an electronic system of density $\rho$ in the chemical potential $\mu$ as
\begin{align}
    E(v,\mu) &= \inf\nolimits_{\rho} \bigl(  \mathcal H(\rho,\mu) + (v \vert \rho) \bigr), \label{eq:LFvrho}\\
    \mathcal H(\rho,\mu) &= \sup\nolimits_{v} \bigl( E(v,\mu) - (v \vert \rho) \bigr), \label{eq:LFrhov}
\end{align}
where $(v \vert \rho) = \int v(\mathbf r) \rho(\mathbf r)\,\mathrm d \mathbf r$. The transformations $E(v,\mu) \leftrightarrow \mathcal E(v,N)$ in Eqs.\,\eqref{eq:LFmuN} and \eqref{eq:LFNmu} and $E(v,\mu) \leftrightarrow \mathcal H(v,N)$ in Eqs.\,\eqref{eq:LFvrho} and~\eqref{eq:LFrhov} are partial skew Legendre--Fenchel transformations, also known as partial skew conjugations. We use the term \emph{partial} to indicate that only one variable is transformed, while \emph{skew} indicates that the transformation changes a convex variable into a concave variable and vice versa. 

We are here mainly interested in the full transformations $\mathcal{E}(v,N) \leftrightarrow \mathcal{H}(\rho,\mu)$, obtained by combining the partial transformations given above as
\begin{align}
    \mathcal E(v,N) &= \sup\nolimits_{\mu} \inf\nolimits_\rho \bigl( \mathcal H(\rho,\mu) + (v \vert \rho)  + \mu N\bigr),  \label{eq:LFE}\\
    \mathcal H(\rho,\mu) &= \inf\nolimits_{N} \sup\nolimits_v \bigl( \mathcal E(v,N) - (v \vert \rho) - \mu N\bigr). \label{eq:LFH}
\end{align}
In particular, we shall use the Hohenberg--Kohn variation principle in Eq.~\eqref{eq:LFE} to calculate the OF-DFT ground-state energy $\mathcal E(v,N)$ of an $N$-electron system in an external potential $v$ from the universal density functional $\mathcal H(\rho,\mu)$ for given chemical potential $\mu$.

\begin{figure}
\includegraphics[width=1.0\columnwidth]{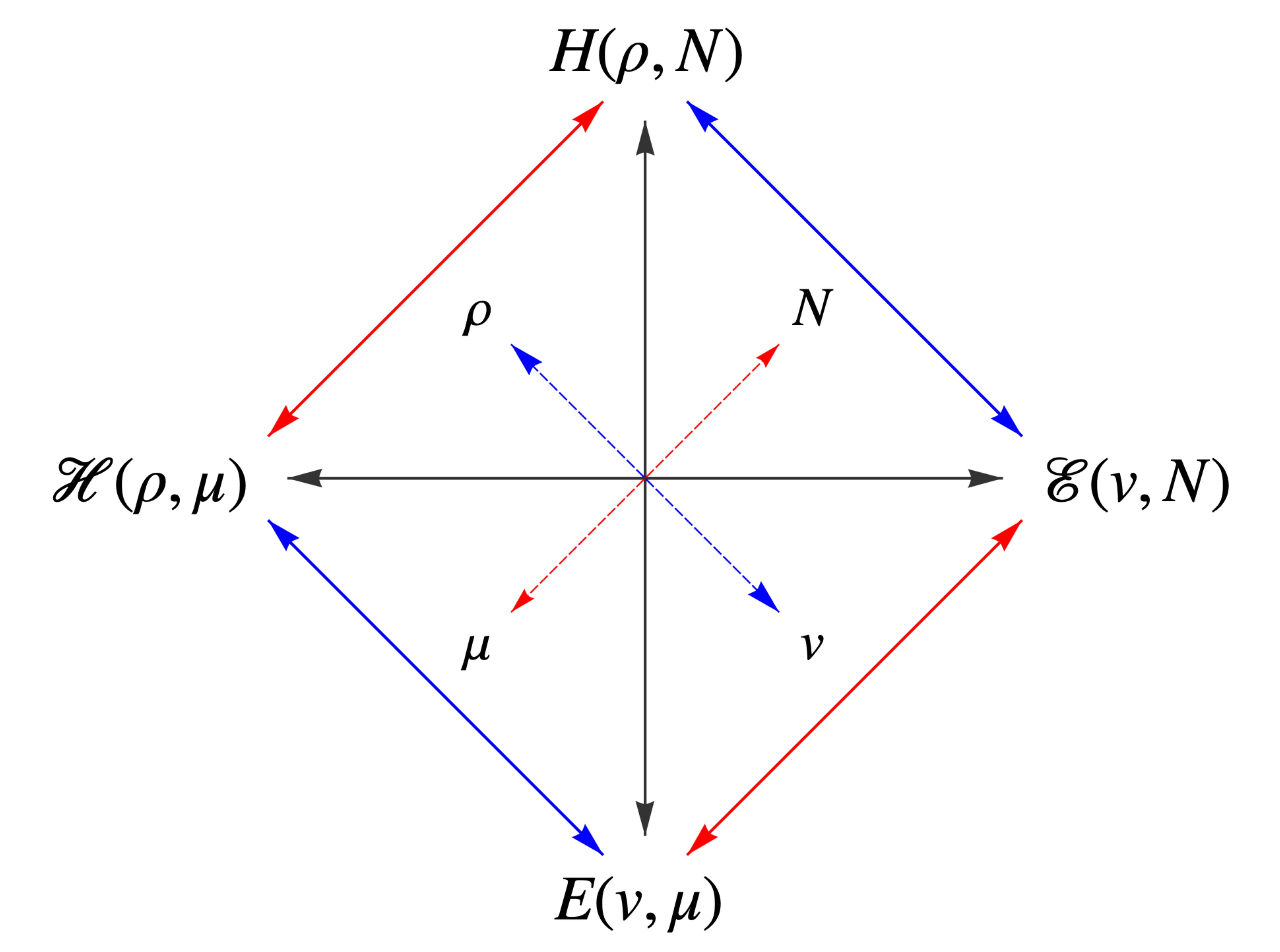}
\caption{The four-way correspondence in OF-DFT. The black horizontal and vertical arrows represent the relationships between each functional by bi-conjugation of both variables simultaneously. The solid blue diagonal arrows indicate skew conjugation of $v \leftrightarrow \rho$, the solid red diagonal arrows indicate skew conjugation of $\mu \leftrightarrow N$. The dashed blue and red arrows indicate the dual relationships between the relevant variables.}\label{fig:FWC}
\end{figure}

The relationship between the three energy functionals introduced above is illustrated in Figure\,\ref{fig:FWC}, which also includes a fourth energy function $H(\rho,N)$, representing the energy for a given density $\rho$ and particle number $N$. It is obtained by the full Legendre--Fenchel transformation of $E(v,\mu)$,
\begin{align}
    E(v,\mu) &= \inf\nolimits_{N}\inf\nolimits_{\rho} \bigl(  H(\rho,N) + (v \vert \rho) - \mu N\bigr), \label{eq:LFEH}\\
    H(\rho,N) &= \sup\nolimits_{\mu}\sup\nolimits_{v} \bigl( E(v,\mu) - (v \vert \rho)+ \mu N\bigr) \label{eq:LFHE}.
\end{align}
As illustrated by arrows in Figure\,\ref{fig:FWC}, the convex--convex function $H(\rho,N)$ is related by partial Legendre--Fenchel transformations to the same saddle functions $\mathcal E(v,N)$ and $\mathcal H(\rho,N)$ as is the concave--concave function $E(v,\mu)$,
\begin{align}
    H(\rho,N) &= \sup\nolimits_{v} \bigl(  \mathcal E(v,N) - (v \vert \rho) \bigr), \label{eq:LFHE1}\\
    \mathcal E(v,N) &= \inf\nolimits_{\rho} \bigl( H(\rho,N) + (v \vert \rho) \bigr) 
\end{align}
and 
\begin{align}
    H(\rho,N) &= \inf\nolimits_{\rho} \bigl(  \mathcal H(\rho,\mu) - \mu N \bigr), \label{eq:LFHE2}\\
    \mathcal H(\rho,\mu) &= \sup\nolimits_{v} \bigl( H(\rho,N) + \mu N  \bigr). \label{eq:LFEH2}
\end{align}
For a proof of this nontrivial point, see Ref.~\citenum{HJO}. Combining the partial transformations of Eqs.~\eqref{eq:LFHE1}--\eqref{eq:LFEH2}, we obtain 
\begin{align}
    \mathcal E(v,N) &= \inf\nolimits_{\mu} \sup\nolimits_\rho \bigl( \mathcal H(\rho,\mu) + (v \vert \rho)  + \mu N\bigr),  \label{eq:LFEx}\\
    \mathcal H(\rho,\mu) &= \sup\nolimits_{N} \inf\nolimits_v \bigl( \mathcal E(v,N) - (v \vert \rho) - \mu N\bigr), \label{eq:LFHx}
\end{align}
which differ from Eq.~\eqref{eq:LFE} and Eq.~\eqref{eq:LFH} only in the order of the minimization and maximization. We thus conclude that the Hohenberg--Kohn variation principle in Eq.~\eqref{eq:LFE} is a minimax saddle-point optimization.

In convex analysis, the four-way relationship $E \leftrightarrow \mathcal H \leftrightarrow H \leftrightarrow \mathcal E \leftrightarrow E$ depicted in Figure\;\ref{fig:FWC} is known is the \emph{four-way correspondence}~\cite{Rockafellar1968}. Since each function in the four-way correspondence can be generated from each of the other three functions, they contain the same information only represented in different ways: in terms of one of the dual variables $\rho$ and $v$ and one of the dual variables $\mu$ and $N$. For an early discussion of the relationship between
the energy functions $E$, $\mathcal H$, $H$ and $\mathcal E$ by analogy with the Maxwell relations, see Ref.~\citenum{nalewajski1982legendre}.

We have previously used the four-way correspondence to illustrate the relationships between the bifunctionals encountered in density-functional theories modified to account for the presence of an external magnetic field~\cite{Reimann2017}. In general, the saddle functions of the four-way correspondence are not unique but consist of equivalence classes of functions; in the special case of the GCE energy functions, the saddle functions are uniquely determined, as shown by Helgaker, J{\o}rgensen and Olsen in Ref.~\citenum{HJO}. 

Having established the Hohenberg--Kohn variation principle for $\mathcal E(v,N)$ as a minimax optimization problem, it is then necessary to identify the universal density functional $\mathcal H(\rho,\mu)$. Returning to the Hohenberg--Kohn variation principle for $E(v,\mu)$ in Eq.\,\eqref{eq:Evmu} and proceeding in the usual Levy--Lieb constrained-search manner~\cite{Levy1979,Lieb1983}, we obtain
\begin{align}
    E(v,\mu) &= \inf_\rho \bigl( \mathcal  H_\text{LL}(\rho,\mu) +  ( v  \vert  \rho) \bigr) 
\end{align}
where we have introduced the Levy--Lieb constrained-search functional
\begin{align}
    \mathcal H_\text{LL}(\rho,\mu) &= \inf_{w_N,\psi_N \mapsto \rho} \sum_N w_N  \bigl\langle\psi_N \left\vert\, T_N + W_N  - \mu \,\right\vert \psi_N \bigr\rangle . \label{eq:HLL}
\end{align}
The question now arises regarding the relationship of $\mathcal H_\text{LL}(\rho,\mu)$ to the density functional $\mathcal H(\rho,\mu)$ in Eq.~(\ref{eq:LFrhov}). Both functional are admissible in the sense that both give the correct energy $E(v,\mu)$ in the Hohenberg--Kohn variation principle. In general, $\mathcal H(\rho,\mu)$ defined as the partial skew conjugate to $E(v,\mu)$ is a lower bound to all admissible density functionals, so $\mathcal H(\rho,\mu) \leq \mathcal H_\text{LL}(\rho,\mu)$. The equality $\mathcal H_\text{LL}(\rho,\mu) = \mathcal H(\rho,\mu)$ then follows by showing that $\mathcal H_\text{LL}(\rho,\mu)$ is lower semi-continuous and convex in $\rho$ for each $\mu$ and upper semi-continuous and concave in $\mu$ for each $\rho$; we do not give the proofs here but refer instead to Ref.~\citenum{HJO}.

Making use of the identification $\mathcal H = \mathcal H_\text{LL}$ and introducing the GCE universal density functional, 
\begin{align}
    \mathcal F(\rho) &= \inf_{w_N,\psi_N \mapsto \rho} \sum_N w_N  \bigl\langle\psi_N \left\vert\, T_N + W_N  \right\vert \psi_N \bigr\rangle ; \label{eq:FL}
\end{align}
we may write the Hohenberg--Kohn variation principle as a minimax problem 
\begin{align}
    \mathcal{E}(v, N) &= \sup_{\mu \in \mathbb{R}} \inf_{\rho \in \chi} \mathcal L(\rho,\mu) = \inf_{\rho \in \chi} \sup_{\mu \in \mathbb R} \mathcal L(\rho,\mu) \label{Esaddle}
\end{align}
where the objective function is given by
\begin{align}
    \mathcal{L}_{v,N}(\rho,\mu) = \mathcal{F}(\rho) + (v - \mu |\rho) + \mu N.
\label{Lfunc}
\end{align}
The optimization of the energy is greatly simplified by the convexity of $\mathcal L_{v,N}(\rho,\mu)$ in $\rho$ and the concavity of $\mathcal L_{v,N}(\rho,\mu)$ in $\mu$: for a given external potential $v$ and particle number $N$, a solution $\mathcal{E}(v,N) = \mathcal L_{v,N}(\rho_\text{min},\mu_\text{max})$ may or may not exist; if it exists, then the solution is unique (but may be degenerate). Moreover, since $\mu$ is a one-dimensional real parameter, the solution is a first-order saddle point, for which robust and efficient optimization schemes exist. 

Introducing $N_\rho = \int \! \rho(\mathbf{r}) \mathrm{d}\mathbf{r}$ and rearranging the objective function $\mathcal L_{v,N}(\rho,\mu)$ slightly, we may write the Hohenberg--Kohn variation principle in the manner
\begin{equation}
    \mathcal{E}(v, N) = \sup_{\mu \in \mathbb{R}} \inf_{\rho \in \chi} \left( \mathcal{F}(\rho) + (v \vert \rho) - \mu (N_\rho - N) \right).
\end{equation}
The Hohenberg--Kohn minimax optimization problem is hence a constrained minimization problem, where we minimize 
$\mathcal{F}(\rho) + (v \vert \rho)$ subject to the constraint that $\rho$ contains $N$ electrons. The objective function $\mathcal L_{v,N}(\rho,\mu)$ is thus a Lagrangian with an undetermined Lagrange multiplier~$\mu$. Previous attempts to optimize the OF-DFT have taken this approach to the optimization of the energy, but without taking into account the convex--concave structure of the Lagrangian and the uniqueness of the saddle-point solution.

Having recognized the simplicity of the OF-DFT minimax optimization problem, it should also be recognized that the exact density functional $\mathcal F(\rho)$ is nondifferentiable; more precisely, it is everywhere discontinuous but everywhere lower semi-continuous. In principle, therefore, we cannot differentiate the density functional and  identify solutions by vanishing gradients. This complication can be overcome by Moreau--Yosida regularization as discussed in Ref.~\citenum{Teale2014}. In the present work, we consider approximations to $\mathcal{H}(\rho, \mu)$ that are simple semi-local density functionals, for which the required derivatives may be readily evaluated.

\subsection{Finite basis-set implementation}\label{sec:FBI}
To perform practical calculations, we optimize the Lagrangian of Eq.\,(\ref{Lfunc}), where we decompose $\mathcal{F}$ in the manner
\begin{equation}\label{eq:Energy}
    \mathcal{F}(\rho) = T_\text{s}(\rho) + E_\text{J}(\rho) + E_\text{xc}(\rho).
\end{equation} 
Here, we follow a similar convention to that used in KS-DFT, where $T_\text{s}(\rho)$ is the noninteracting kinetic-energy functional, $E_\text{J}(\rho)$ is the classical Coulomb energy, and $E_\text{xc}(\rho)$ is the exchange--correlation energy. Other decompositions are equally valid but this choice means that the wide variety of approximations developed for $E_\text{xc}(\rho)$ in KS-DFT may be employed in this context, provided appropriate approximate functionals for $T_\text{s}(\rho)$ can be derived. The expressions for the classical Coulomb repulsion energy
\begin{equation}
    E_\text{J}(\rho) = \int \int \frac{\rho(\mathbf{r})\rho(\mathbf{r}^\prime)}{|\mathbf{r} - \mathbf{r}^\prime|} \mathrm{d}\mathbf{r} \mathrm{d}\mathbf{r}^\prime
\end{equation}
and electron--nuclear attraction energy
\begin{equation}
    E_v(\rho) = \int\!\! \rho(\mathbf{r}) v(\mathbf{r}) \mathrm{d}\mathbf{r}
\end{equation}
can be evaluated exactly. Unlike in KS-DFT, two contributions to the energy must then be approximated -- namely, the noninteracting kinetic  energy $T_\text{s}(\rho)$ and the exchange--correlation energy $E_\text{xc}(\rho)$. A wide range of approximations are available for the exchange--correlation energy~\cite{mardirossian2017thirty}; all of those at the local-density and generalized-gradient levels of approximation can be readily applied in OF-DFT. For $T_\text{s}(\rho)$, many approximations have been suggested~\cite{semilocal} but such approximations are far fewer in number than those available for $E_\text{xc}(\rho)$. Furthermore, in many cases their accuracy in self-consistent calculations is also unclear as their development and testing has often been carried out based on fixed input densities.

To ensure that the electron density remains positive everywhere during the optimization, we represent its square root as
\begin{equation}\label{rho_exp}
    \sqrt{\rho(\mathbf{r})} = \sum_p c_p \varphi_p(\mathbf{r}),
\end{equation}
where $c_p$ are expansion coefficients to be determined in the optimization and $\varphi_p(\mathbf{r})$ are a set of Gaussian basis functions. To perform the optimization of the Lagrangian defined in Eq.\,(\ref{Lfunc}) with respect to the electron density, we evaluate its partial derivative with respect to the electron density to yield the Euler--Lagrange equation,
\begin{align}
    \pdv{\mathcal L_{v,N}(\rho,\mu)}{\rho(\mathbf{r})} &= \fdv{T_\text{s}(\rho)}{\rho(\mathbf{r})} + \fdv{E_v(\rho)}{\rho(\mathbf{r})}  \nonumber \\ 
    &\qquad+ \fdv{E_\text{J}(\rho)}{\rho(\mathbf{r})} + \fdv{E_\text{xc}(\rho)}{\rho(\mathbf{r})} - \mu = 0.
\label{eq:Euler}
\end{align}
Similarly, the optimization of the chemical potential can be carried out using the partial derivative,
\begin{equation}
    \pdv{\mathcal L_{v,N}(\rho,\mu)}{\mu} = N - N_\rho,
\label{eq:muG}
\end{equation}
which is simply the error in the particle number $N_\rho$ at a given step of the optimization relative to the target particle number $N$. 

Introducing the finite basis-set expansion of Eq.\,(\ref{rho_exp}), the gradient with respect to the expansion coefficients $c_p$ can be evaluated as
\begin{align}\label{eq:grad_bas}
&\pdv{\mathcal L_{v,N}(\rho,\mu)}{c_p} = \nonumber \\ & \quad \left \langle \varphi_p \left | \fdv{T_\text{s}(\rho)}{\rho(\mathbf{r})} + v(\mathbf{r}) + v_\text{J}(\mathbf{r}) + v_\text{xc}(\mathbf{r}) - \mu \right | \sqrt{\rho} \right \rangle
\end{align}
and the gradient with respect to the chemical potential as
\begin{equation}\label{eq:mu_grad_bas}
    \pdv{\mathcal L_{v,N}(\rho,\mu)}{\mu} = N - \int \!\! \rho(\mathbf{r}) \mathrm{d}\mathbf{r}.
\end{equation}
The Hessian with respect to the expansion coefficients is given by
\begin{equation}
    \begin{split}
        \pdv{\mathcal L_{v,N}(\rho,\mu)}{c_p}{c_q} &= \pdv{T_\text{s}(\rho)}{c_p}{c_q}  + \pdv{E_v(\rho)}{c_p}{c_q} + \pdv{E_\text{J}(\rho)}{c_p}{c_q}\\
        & + \pdv{E_\text{xc}(\rho)}{c_p}{c_q} - 2\mu \langle \varphi_p | \varphi_q \rangle   
    \end{split}\label{eq:Hess}
\end{equation}
where 
\begin{align}
    \pdv{E_v(\rho)}{c_p}{c_q} &= 2  \langle \varphi_p | v(\mathbf{r}) | \varphi_q \rangle \\
    \pdv{E_\text{J}(\rho)}{c_p}{c_q} &= 2\sum_{rs} c_r c_s \qty[ 2 \langle pq|rs \rangle + \langle pr|qs \rangle].
\end{align}
In the present work, we consider functionals of LDA and GGA type for $T_\text{s}(\rho)$ and $E_\text{xc}(\rho)$, which may be expressed in the general form
\begin{equation}
    F(\rho) = \int \!\! f(\rho,\grad{\rho})\,\mathrm{d}\mathbf{r},
\end{equation}
where $f$ is the integrand, depending on the density and its gradient, the contributions to the Hessian Eq.~\eqref{eq:Hess} of which take the form
\begin{equation}
    \begin{split}
        \pdv{F(\rho)}{c_p}{c_q} =& \left \langle \varphi_p \left| \pdv{\rho}{c_p}{c_q}\pdv{f}{\rho} +  \pdv[2]{f}{\rho}\pdv{\rho}{c_p}\pdv{\rho}{c_q}  \right | \varphi_q \right \rangle\\ 
        +&\left \langle \varphi_p \left|  \pdv{\grad\rho}{c_p}{c_q}\pdv{f}{\grad\rho}+ \pdv[2]{f}{\grad\rho} \pdv{\grad\rho}{c_p}\pdv{\grad\rho}{c_q} \right | \varphi_q \right \rangle \\
        +&\left \langle \varphi_p \left| \pdv{f}{\grad\rho}{\rho}\qty[\pdv{\rho}{c_q}\pdv{\grad\rho}{c_p} + \pdv{\rho}{c_p}\pdv{\grad\rho}{c_q}] \right | \varphi_q \right \rangle 
    \end{split}
\end{equation}
where only the first term is present for LDA type functionals. 

The mixed Hessian contributions arising from the density expansion and chemical potential have the simpler form
\begin{equation}
    \pdv{\mathcal L_{v,N}(\rho,\mu)}{\mu}{c_p} = -2\langle \varphi_p | \sqrt{\rho} \rangle,
\end{equation}
while those arising purely from the chemical potential vanish,
\begin{equation}\label{mu_mu_der}
    \frac{\partial^2\mathcal L_{v,N}(\rho,\mu)}{\partial \mu^2} = 0.      
\end{equation}
The Lagrangian given in Eq.\,(\ref{Lfunc}) and its derivatives in Eqs.\,(\ref{eq:grad_bas})--(\ref{mu_mu_der}) represent the objective function and all derivative information required to employ direct optimization methods up to second-order in finite basis sets.

\subsection{Optimization methods in OF-DFT}\label{subsec:opt}
\label{sec:optmethods}
Several approaches have been put forward to carry out the constrained optimization required for OF-DFT calculations. In the present work, we have implemented two of these for comparison with a new trust-region second-order saddle-point method. We now give a brief overview of the three methods we have considered in this work.  

\subsubsection{The Levy--Perdew--Sahni/Lopez-Acevedo method}
In 1984, Levy, Perdew and Sahni~\cite{levy1984exact} (LPS) showed that the OF-DFT optimization could be cast in a form amenable to solution by any standard Kohn--Sham program. Their idea was to express the Lagrangian in terms of the vW kinetic energy $T_\text{vW}(\rho)$ and a residual Pauli kinetic energy $T_\theta(\rho) = T_\text{s}(\rho) - T_\text{vW}(\rho)$ in the manner,
\begin{align}
    \mathcal{L}_{v,N}(\rho,\mu) &= T_\text{vW}(\rho) + T_\theta(\rho) + E_v(\rho) \nonumber \\
    &\quad + E_\text{J}(\rho)+ E_\text{xc}(\rho) - \mu (N_\rho - N).
\end{align}
The first two terms account for the chosen $T_\text{s}(\rho)$, whilst the remaining contributions are the same as those from standard Kohn--Sham theory. The Euler--Lagrange equation then becomes,
\begin{align}
    \fdv{\mathcal L_{v,N}(\rho,\mu)}{\rho(\mathbf{r})} &= \fdv{T_\text{vW}(\rho)}{\rho(\mathbf{r})} +  \fdv{T_\theta(\rho)}{\rho(\mathbf{r})} + \fdv{E_v(\rho)}{\rho(\mathbf{r})}  
    \nonumber \\ & \quad + \fdv{E_\text{J}(\rho)}{\rho(\mathbf{r})} + \fdv{E_\text{xc}(\rho)}{\rho(\mathbf{r})} - \mu = 0.
\end{align}
Introducing the effective potential,
\begin{align}
    v_\text{eff}(\mathbf{r}) = \fdv{T_\theta(\rho)}{\rho(\mathbf{r})} + \fdv{E_v(\rho)}{\rho(\mathbf{r})}  + \fdv{E_\text{J}(\rho)}{\rho(\mathbf{r})} + \fdv{E_\text{xc}(\rho)}{\rho(\mathbf{r})},
\end{align}
evaluating the functional derivative of the vW kinetic energy 
\begin{equation}
    T_\text{vW}(\rho) = \int \!\!\sqrt{\rho(\mathbf{r})} \left(-\frac{1}{2} \nabla^2\right) \sqrt{\rho(\mathbf{r})} \mathrm{d} \mathbf{r},
\end{equation}
multiplying through by $\sqrt{\rho(\mathbf{r})}$ and rearranging, we arrive at the Kohn--Sham-like equation,
\begin{align}\label{KS_OFDFT}
    \left[ -\frac{1}{2}\nabla^2 + v_\text{eff}(\mathbf{r}) \right] \sqrt{\rho(\mathbf{r})} = \mu \sqrt{\rho(\mathbf{r})}.
\end{align}
Here $\sqrt{\rho(\mathbf{r})}$ is analogous to a single orbital, $\mu$ is its associated eigenvalue, and $v_\text{eff}(\mathbf r)$ contains the functional derivative of $T_\theta(\rho)$ with respect to the electron density in addition to the terms usually found in the Kohn--Sham effective potential. This observation led LPS to assert that the OF-DFT problem could be solved easily by any standard Kohn--Sham program~\cite{levy1984exact}.

In practice, the ease with which the solution of Eq.\,(\ref{KS_OFDFT}) may be obtained is strongly dependent on the choice of approximate $T_\text{s}(\rho)$. It has been observed that solutions for functionals of the form $T_\text{s}(\rho) = \gamma T_\text{TF}(\rho) + \lambda T_\text{vW}(\rho)$, where $T_\text{TF}(\rho)$ is the Thomas--Fermi kinetic energy, may be obtained when $\gamma = 1$ and $\lambda = 1$. However, for other values of $\gamma$ and $\lambda$ and for functionals outside the TF--vW family,  significant convergence issues are encountered. Even with $\gamma = \lambda = 1$, convergence can be slow, standard self-consistent-field (SCF) acceleration techniques fail and strong Pulay damping must be employed, leading to a large number of iterations -- see, for example, Refs.~\citenum{Karasiev2012,lehtomaki2014orbital}. 

In 2014, Lopez-Acevedo (LA) and co-workers~\cite{lehtomaki2014orbital} proposed a solution to these convergence problems for members of the $\gamma$TF$\lambda$vW family of functionals. For this family of functionals,  $T_\theta(\rho) 
=  \gamma T_\text{TF}(\rho) + (\lambda - 1) T_\text{vW}(\rho)$. Adding the second term to the kinetic-energy operator and dividing by $\lambda$, we obtain
\begin{align}\label{KS_OFDFT_LA}
    \left[ -\frac{1}{2}\nabla^2 + \frac{1}{\lambda} v^\prime_\text{eff}(\mathbf{r}) \right] \sqrt{\rho(\mathbf{r})} = \frac{\mu}{\lambda} \sqrt{\rho(\mathbf{r})}.
\end{align}
where $v^\prime_\text{eff}(\mathbf r) = v_\text{eff}(\mathbf r) - (\lambda -1)T_\text{vW}(\rho)$. With this modification, self-consistent OF-DFT solutions can be obtained with a range of $\gamma$ and $\lambda$ values. 

There are two important limitations to this approach; firstly, the slow convergence and requirement for heavy damping of the SCF iterations remains;  secondly, the approach is only effective for the $\gamma$TF$\lambda$vW family of functionals. The first limitation may not be severe -- in the original work by LA and co-workers, for example, this technique was used with the projector-augmented-wave (PAW) method. Although it was observed that 10--100 more iterations are required for OF-DFT solutions of atoms compared with KS-DFT, this slow convergence affects only the initial setup phase of the calculations for each atom and not the subsequent solution of the entire system. The restriction to $\gamma$TF$\lambda$vW-type functionals is more severe, however, limiting the usefulness of this method for studying general approximations to $T_\text{s}(\rho)$.

\subsubsection{The Cohen--Chan--Handy (CCH) method}
A more general approach was put forward by Cohen, Chan and Handy~\cite{Chan2001} (CCH) in 2001. The context for this approach is a direct optimization technique in a Gaussian basis set using the expansion and some of the derivatives introduced previously. The optimization proceeds in two phases. In the first phase, $\mathcal{L}_{v,N}(\rho,\mu)$ is optimized with respect to $\rho$ for a fixed trial value of the chemical potential $\mu$, using only the density gradient and (optionally) Hessian terms in Eqs.\,(\ref{eq:grad_bas}) and (\ref{eq:Hess}). The electron-number error (the gradient with respect to the chemical potential) is then determined and the chemical potential adjusted so as to step in the direction to reduce this error. The Lagrangian $\mathcal{L}_{v,N}(\rho,\mu)$ is then re-optimized with respect to $\rho$ at this new value of $\mu$ and the electron-number error recalculated.  This process is repeated until the electron-number error changes sign. 
At this point, the optimization moves to the second phase, in which a bisection is performed to precisely identify the value of $\mu$ for which the particle number error is zero. At each step of the bisection, a new value of $\mu$ is chosen and $\mathcal{L}_{v,N}(\rho,\mu)$ is again optimized for this fixed $\mu$ until the bisection is complete. 

The CCH optimization can be performed using only first-order information, with the gradient from Eq.~(\ref{eq:grad_bas}) and utilizing a Hessian update scheme for the convergence of $\mathcal{L}_{v,N}(\rho,\mu)$ at each step in the two phases. Alternatively, the exact Hessian with respect to the density can be calculated and used in a second-order approach, giving a more robust and rapid convergence. 

The CCH scheme can be applied with any semi-local approximation to $T_\text{s}(\rho)$. However, in Ref.~\citenum{Chan2001}, only results from $\gamma$TF$\lambda$vW functionals were reported. In their paper, CCH remark that \emph{``Due to the highly nonquadratic nature of the kinetic energy, the optimization ... is a nontrivial problem. The iterative self-consistent procedure commonly used in Kohn–Sham calculations does not work, and we require more robust minimization techniques.''} ... \emph{``first derivative methods such as conjugate gradient minimization and quasi-Newton search perform poorly, requiring many hundreds of iterations to achieve convergence''} ... \emph{``these problems are exacerbated if we incorporate the constraint $\int \rho(\mathbf{r}) \mathrm{d} \mathbf{r} = N$ directly into an energy minimization''}. 

\subsubsection{The trust-region image method (TRIM)}\label{subsec:TRIM}
In the present work, we introduce a flexible framework for all-electron OF-DFT calculations using second-order information, with focus on the robustness and precision of the optimization. To this end, we note the saddle nature of the optimization in Eq.~(\ref{Esaddle}) and that, for the exact $T_\text{s}(\rho)$ and $E_\text{xc}(\rho)$, this functional has only one global first-order saddle point. With this information in mind, we consider optimization of this saddle problem directly, for which the gradient of dimension $M+1$ has the structure
\begin{equation}\label{full_grad}
    \mathbf{g} = \qty(\cdots ,\pdv{\mathcal{L}}{c_p}, \cdots, \pdv{\mathcal{L}}{\mu}),
\end{equation}
while the Hessian of dimension $(M+1) \times (M+1)$ has the structure
\begin{equation}\label{full_hess}
    \mathbf{H} = \mqty(\vdots& \vdots & \vdots&\vdots\\ \hdots & \pdv{\mathcal{L}}{c_p}{c_q} & \hdots & \pdv{\mathcal{L}}{c_p}{\mu} \\\vdots& \vdots & \ddots&\vdots\\ \hdots & \pdv{\mathcal{L}}{\mu}{c_q}& \hdots & 0 ),
\end{equation}
where $M$ is the number of basis functions in which the square root of the density is expanded. The elements of the gradient and Hessian are described in the Theoretical Methods section. 

A simple approach for saddle-point optimizations is the trust-region image method (TRIM), proposed by Helgaker in 1991 in the context of transition-state geometry optimization~\cite{helgaker1991transition}. In this method, the concept of an image function is used~\cite{smith1990find}, defined by considering the image surface as the function having the same gradient and Hessian as the original surface except with the sign of one Hessian eigenvalue and the corresponding gradient element reversed. This simple change means that first-order saddle points in the objective function are transformed into local minima in the image function, guaranteeing local convergence. We note that the TRIM scheme is readily generalized to searches for higher-order transition states, as shown in recent work by Field-Theodore and Taylor~\cite{ALTRUISM}. 

In the present context, the last element of Eq.\,(\ref{full_grad}) corresponding to the chemical potential would have its sign reversed and the corresponding eigenvalue of the Hessian in Eq.\,(\ref{full_hess}) would also have its sign reversed. Whilst not all functions have a corresponding image~\cite{Sun}, in second-order optimization schemes we only require that the second-order Taylor expansion around a given reference point has an image function. Denoting the density expansion coefficients and chemical potential (the optimization variables) collectively by $\mathbf{x} = \{ c_1, \dots, c_n, \mu \}$, the second-order expansion of the Lagrangian around a given point in the optimization $\mathbf{x}_0$ can be written as
\begin{align}
    \mathcal{L}(\mathbf{x}) &= \mathcal{L}(\mathbf{x}_0) + \sum_p g_p (x_p - x_{0,p}) \nonumber \\ &+ \frac{1}{2} \sum_{pq} H_{pq} (x_p-x_{0,p})(x_q-x_{0,q})
\end{align}            
where $g_p$ and $H_{pq}$ are the elements of the gradient and Hessian in Eqs.~(\ref{full_grad}) and (\ref{full_hess}).  

In the TRIM approach, a standard trust-region algorithm is employed for optimization of $\mathcal{L}(\rho,\mu)$. The TRIM step is typically determined in the diagonal representation, where the Hessian is diagonalized by an orthogonal transformation, 
\begin{align}
    \mathbf{h} = \mathbf{U}^\mathrm{T} \mathbf{H} \mathbf{U} 
\end{align}
where $\mathbf{h}$ contains the eigenvalues of $\mathbf{H}$ sorted into ascending order. The gradient is then transformed to this representation via,
\begin{align}
    \mathbf{f}^\mathrm{T} = \mathbf{g}^\mathrm{T} \mathbf{U}.
\end{align}
The signs of the relevant elements of the gradient and Hessian in the diagonal representation are then inverted. In the present case,
\begin{align}
\tilde{f}_1 = -f_1; \hspace{0.1in}  \tilde{f}_p = f_p \hspace{0.1in} (p >1)
\end{align}
and
\begin{align}
\tilde{h}_1 = -h_1; \hspace{0.1in}  \tilde{h}_p = h_p \hspace{0.1in} (p >1).
\end{align}
The TRIM algorithm for determining a step $\mathbf{s}$ is then first to compute the Newton step,
\begin{align}
    s_i = - \tilde{f}_i / \tilde{h}_i.
\end{align}
If $||\mathbf{s}|| \leq t$, where $t$ is the trust radius at a given step, then the Newton step is taken. The trust region is subsequently updated according to the usual rules for trust-region optimization~\cite{Numopt}. Alternatively, if $||\mathbf{s}|| > t$, then a step is instead taken to the boundary of the trust region, according to 
\begin{align}
    s_i = -\tilde{f}_i / (\tilde{h}_i + \lambda)
\end{align}
where $\lambda$ is determined by solving
\begin{align}
    \left[  \sum_i \left(   \frac{\tilde{f}_i}{\tilde{h}_i + \lambda}  \right)^2   \right]^{\frac{1}{2}} = t.
\end{align}
This may be achieved by simple bisection at low computational cost. In principle, the solution of this equation may encounter difficulties, see Refs.~\citenum{ALTRUISM,BB} for further discussion. In practice, we have not observed difficulties with obtaining suitable solutions for $\lambda$.

From the discussion of the four-way correspondence in OF-DFT, we recall that the convex--concave saddle function $\mathcal{L}_{v, N}(\rho,\mu)$ has one global stationary point when the exact $T_\text{s}(\rho)$ and $E_\text{xc}(\rho)$ are employed, although this is not guaranteed for approximate functionals. Second-order methods are well suited to such problems and we shall find that TRIM gives robust and rapid convergence, though the obvious drawback to this is the requirement of a diagonal representation for the Hessian. In the present work, we focus on providing a robust approach to converge all-electron calculations in this context for modestly sized systems, noting that diagonalization may be eliminated with further development for large-scale applications.

\section{Computational Details}\label{sec:compdet}
We have implemented the three optimization methods described previously into our in-house program \textsc{Quest}~\cite{QUEST}. To evaluate the partial derivatives of the semi-local approximations used for $T_\text{s}(\rho)$ and $E_\text{xc}(\rho)$, required for construction of the gradients and Hessians with respect to the expansion coefficients, we use the XCFun package~\cite{XCFun}. The XCFun package uses automatic differentiation to evaluate derivatives of semi-local density-functionals to arbitrary order. 
We have implemented a large number of approximations for $T_\text{s}(\rho)$ into the XCFun package, including all of those in Ref.~\citenum{semilocal}. 

In the present work, we will primarily demonstrate the convergence properties of the optimization methods considered, restricting our attention to a limited selection of density functionals for $T_\text{s}(\rho)$, which includes functionals up to GGA level. For the exchange--correlation term, we utilize the Dirac exchange functional throughout, facilitating comparison with previous works. We have tested the use of other exchange--correlation functionals such as
the PBE functional.\cite{Perdew1996} However, 
since the errors in the kinetic-energy functionals dominate the total errors in the present calculations, use of these
alternative exchange--correlation functionals does not change the conclusions 
regarding quality of the $T_\text{s}(\rho)$ approximations.

For atomic systems with the Born--Oppenheimer electronic Hamiltonian, OF-DFT solutions are spherical for atoms of any nuclear charge $Z$, in contrast to those obtained from spin-polarized KS-DFT calculations. As a consequence, atomic solutions are well represented in basis sets containing only s--type Gaussian functions. Unless otherwise stated, we use a large even-tempered basis set with exponents $3^n$ with $n = -6, \dots, 12$, as used by CCH~\cite{Chan2001}. For molecules, we use basis sets constructed from the primitive exponents of the def2-qzvp basis~\cite{qzvp}. This basis set represents a flexible choice suitable 
for both OF-DFT and KS-DFT calculations, enabling comparison between results from both. We note that, for OF-DFT, it is possible to remove many of the 
higher angular-momenta functions without a significant effect on accuracy -- however, the careful optimization of basis sets for $\sqrt{\rho(\mathbf{r})}$ is left to future work. Here. we consider a set of 28 small molecules at the geometries given in Ref.~\citenum{TealeNMR}.

Although computational scaling is not our primary concern, we have implemented two approaches to reduce the cost of evaluating the required two-electron integrals. For first-order optimizations, the J-engine approach of Alml\"of~\cite{RezaAhmadi1995} is an effective method for accelerating the energy and gradient evaluations, requiring no additional approximation. For second-order optimizations, the exchange--type integrals required in the Hessian construction are most effectively accelerated using resolution-of-identity (RI) techniques. For generation of the auxiliary basis set, we have implemented the AutoAux scheme outlined by Neese~\cite{Stoychev2017}. This approach is based on the product basis and is very conservative; in our preliminary tests, we observed that the energy is accurate to better than $10^{-4}E_\text{h}$ compared with the conventional evaluation. We have employed this procedure throughout, reporting energies to $10^{-4}E_\text{h}$. 

For the direct optimization schemes, we employ several convergence criteria, all of which must be satisfied for iterations to be terminated (here given in
atomic units): 
$||\mathbf{g}|| < 10^{-6}$, $||\mathbf{s}|| < 10^{-5}$, change in the gradient norm between iterations less than $10^{-6}$,
and change in the Lagrangian and chemical potential between iterations 
less than $5\times10^{-6}$. For the LA approach, the iterations are terminated when the change in the energy and the chemical potential are both less than $10^{-6}$.

\section{Results and Discussion}\label{sec:res}
In this section, we discuss the performance of the TRIM optimization scheme. We commence by comparing 
TRIM with the LA and CCH optimization schemes. The rate of convergence for the different approaches is examined, along with the precision with which convergence is achieved. Convergence towards the basis set limit is also considered. We then apply TRIM to atomic systems, examining the convergence properties for different functionals and paying particular attention to the uniqueness of the solutions obtained. Finally, having established the robustness of TRIM for atomic systems we apply it to a selection of molecular systems, where we highlight the importance of the initial guess.


\subsection{Convergence properties}\label{subsec:conv}
For any finite basis-set method, it is desirable (i) that the employed optimization methods are robustly and rapidly convergent in a given basis and 
(ii) that convergence towards the complete basis-set limit is possible, and ideally smooth and rapid. In the following, 
we discuss each of these aspects in turn. It has been noted by Karasiev and Trickey that 
\emph{``An interesting feature of the literature... is that there are more tests of approximate functionals using inputs from other sources (e.g. conventional KS calculations, Hartree--Fock calculations etc.) 
than tests by solving the Euler equation ... A side-effect is that comparatively little is known about the difficulty of solving that equation with approximations other than that of the Thomas--Fermi kind ... and about the relative effectiveness of various solution techniques''}. Here, we aim to address this point and present applications of TRIM to OF-DFT calculations on atoms and molecules.

\subsubsection{Optimization method}
\begin{figure*}
\includegraphics[width=0.5\textwidth]{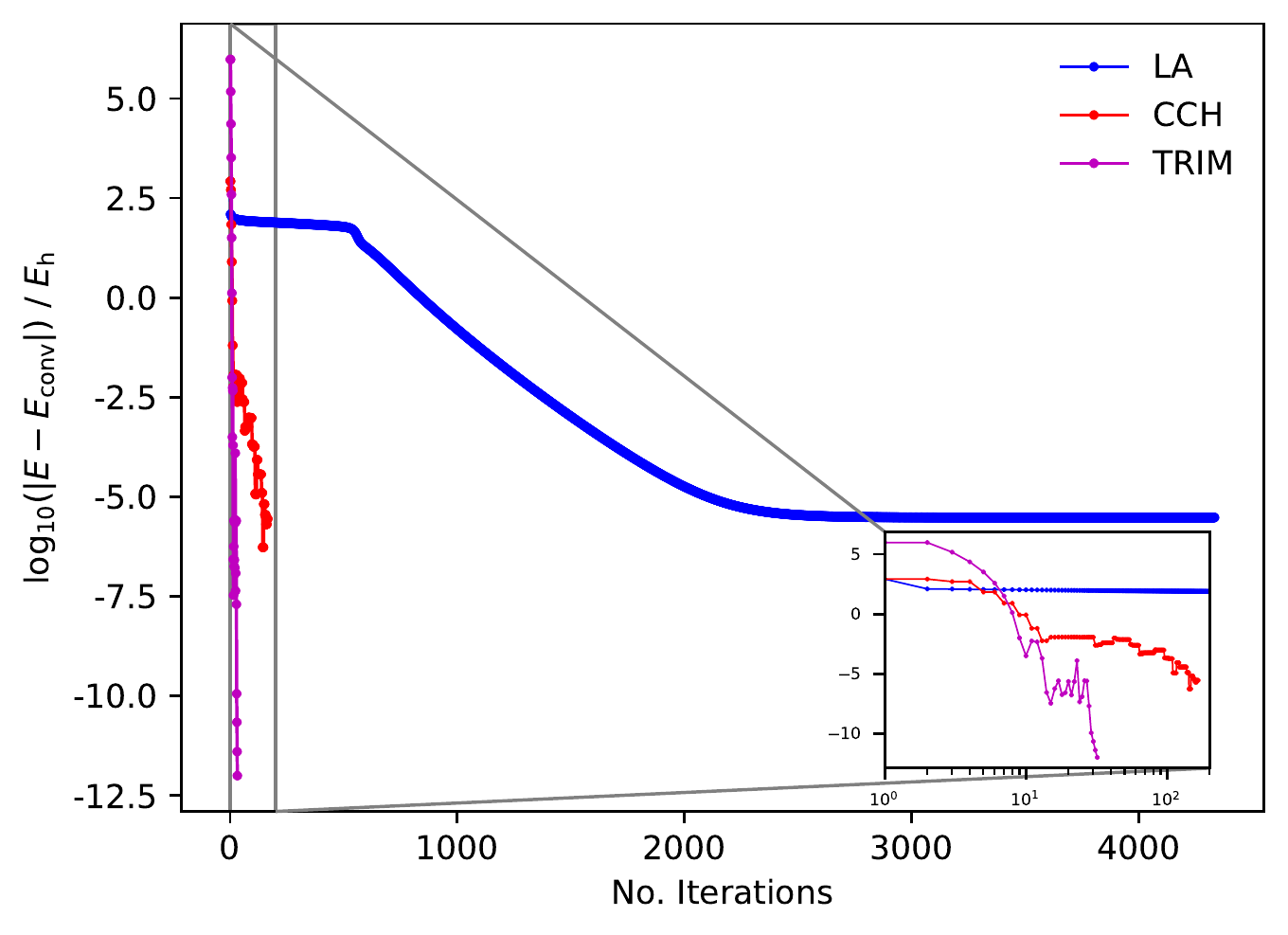}\includegraphics[width=0.5\textwidth]{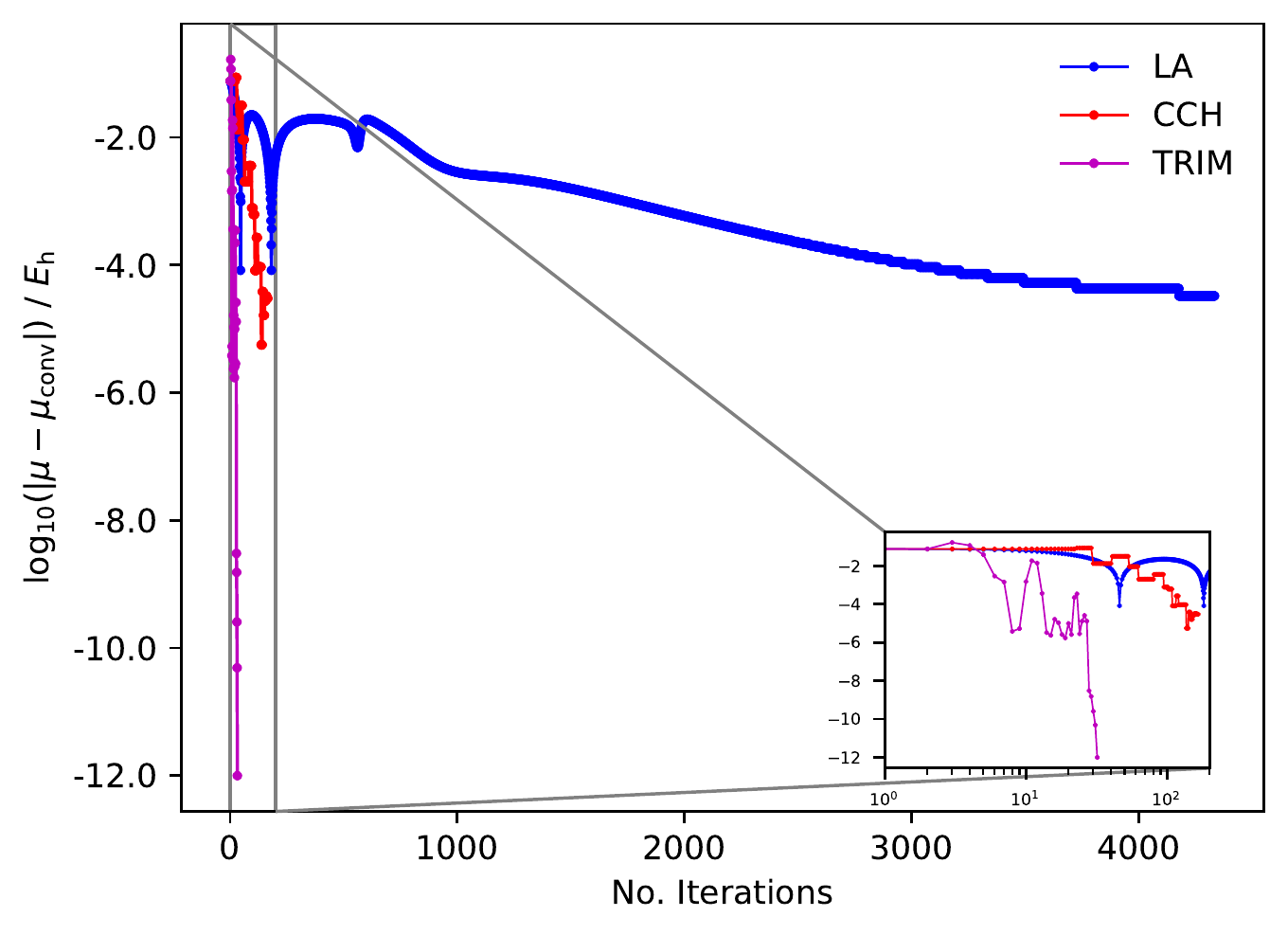}
\caption{Convergence of the LA, CCH and TRIM approaches. In the left panel, the convergence of the energy is plotted as $\log_{10}(|E-E_\text{conv}|)$ 
against the number of energy and gradient / pseudo KS-matrix evaluations. In the right panel, the convergence of the chemical potential is plotted as  $\log_{10}(|\mu-\mu_\text{conv}|)$ against the number of iterations.}\label{fig:conv}
\end{figure*}

The LA scheme was developed to enable convergence for more general members of the $\gamma$TF$\lambda$vW family of functionals. In particular, the values $\gamma=1$ and $\lambda=\frac{1}{5}$ have been identified as giving the most accurate energies of atomic systems. In the remainder of this work,
we denote specific functionals from this family by $\gamma$TF$\lambda$vW($\gamma$,$\lambda$). In Figure\,\ref{fig:conv}, 
we compare the convergence of the LA, CCH and TRIM methods with this functional for the neon atom. Convergence of the energy and the chemical potential are plotted as a function of the number of iterations. 

It is clear from Figure\,\ref{fig:conv} that the LA scheme, whilst convergent, requires a far greater number of iterations than the CCH or TRIM schemes -- here taking 4334 iterations. As described previously, 
this slow convergence is due to the significant damping required to converge the SCF equations. These observations are in line with the original implementation in Ref.\,\citenum{lehtomaki2014orbital}. We use a single damping factor that is adequate to converge all of the $\gamma$TF$\lambda$vW family of functionals considered here, when applied to atoms with $Z \leq 18$. We observe that an increasingly stronger damping is required as $\lambda$ decreases and/or as $Z$ increases. In addition, our convergence criteria are intentionally conservative to ensure that the chemical potential is well converged. As such, optimization of the damping parameter for a given atom or functional may reduce the number of iterations. However, for general application -- 
particularly as a starting point for molecular calculations, which may contain multiple atoms of different $Z$ -- a conservative choice of damping parameter as employed in this work is essential. 

The CCH scheme is more rapidly convergent than the LA scheme, both in its gradient-only form and when the Hessian of Eq.\,(\ref{eq:Hess}) is used in the sub-optimizations of $\mathcal{L}(\rho,\mu)$ for fixed $\mu$. Nonetheless, the nested nature of the optimization means that, whilst the convergence is robust, it is not very rapid, with the the second-order approach requiring 166 energy, gradient and Hessian evaluations. The terraced nature of the plot reflects the bracketing and bisection phases of the optimization. Each terrace becomes shorter (fewer iterations) as the restarting becomes progressively more effective closer to convergence. 

By contrast, the TRIM scheme exhibits very rapid convergence, with quadratic convergence in the last few iterations, as would be expected of a second-order method. Moreover, convergence is achieved in just 32 iterations. Compared with the LA and CCH methods, the convergence of the TRIM method with respect to the energy and chemical potential is very precise, reaching a limit far exceeding the required tolerance, due to the quadratic convergence in the final steps.   

This pattern of convergence is typical across the atomic systems and functionals considered in this work. The mean, median and maximum number of TRIM iterations required for convergence with each functional for the atoms with $Z \leq 18$ are shown in Table~\ref{tab:atom_conv}. One may expect the convergence of $\gamma$TF$\lambda$vW functionals to become more difficult as $\lambda$ decreases; however, the performance of TRIM is remarkably consistent across this family of functionals. Again, it is clear that the TRIM approach offers rapid convergence for most of the functionals tested, including the OL1~\cite{OL1/OL2}, P92~\cite{P92} and E00~\cite{E00} GGAs. 

Conjoint kinetic-energy functionals are functionals in which the enhancement factors take the same form as those for the corresponding exchange--correlation functionals, but re-parameterized for the kinetic energy~\cite{Lee1991}. 
Interestingly, the conjoint functionals conjB86a and conjPW91 are more difficult to converge than the  other functionals -- in particular, for the hydrogen and helium atoms, as shown by the maximum number of iterations. We shall see later that these functionals are not recommended for general use in self-consistent calculations for other reasons. 
\begin{table*}
\caption{Convergence statistics for TRIM on the atoms with $1 \leq Z \leq 18$. The maximum, minimum, mean and median number of iterations required for convergence are given for each functional.}\label{tab:atom_conv}
\begin{tabular}{lrrrrrrrrr}
\hline
\hline
	&$\gamma$TF$\lambda$vW 	&$\gamma$TF$\lambda$vW  &$\gamma$TF$\lambda$vW 	& $\gamma$TF$\lambda$vW &OL1	&P92	&E00	& conjB86a	& conjPW91 \\
$\gamma$	& $1.0$  	&$1.0$     & $1.0$    & $0.697$ & 	&	&	&	&  \\
$\lambda$ 	& $1/9$	        &$0.186$     &  $1/5$	& $0.599$	&	&	&	&	&  \\
\hline
Max	         &39	& 50	&32	&11	& 45	&39 &14	&165&	76 \\
Min	         &9	&9	&9	&8	&9	&9	&10	&11	&14 \\
Mean	&19	&17	&15	&10	&17	&16	&12	&30	&29\\
Median	&13	&13	&13	&11	&13	&13	&12	&16	&20 \\
\hline
\hline
\end{tabular}
\end{table*}

\subsubsection{Basis set}
\begin{figure}
\includegraphics[width=\columnwidth]{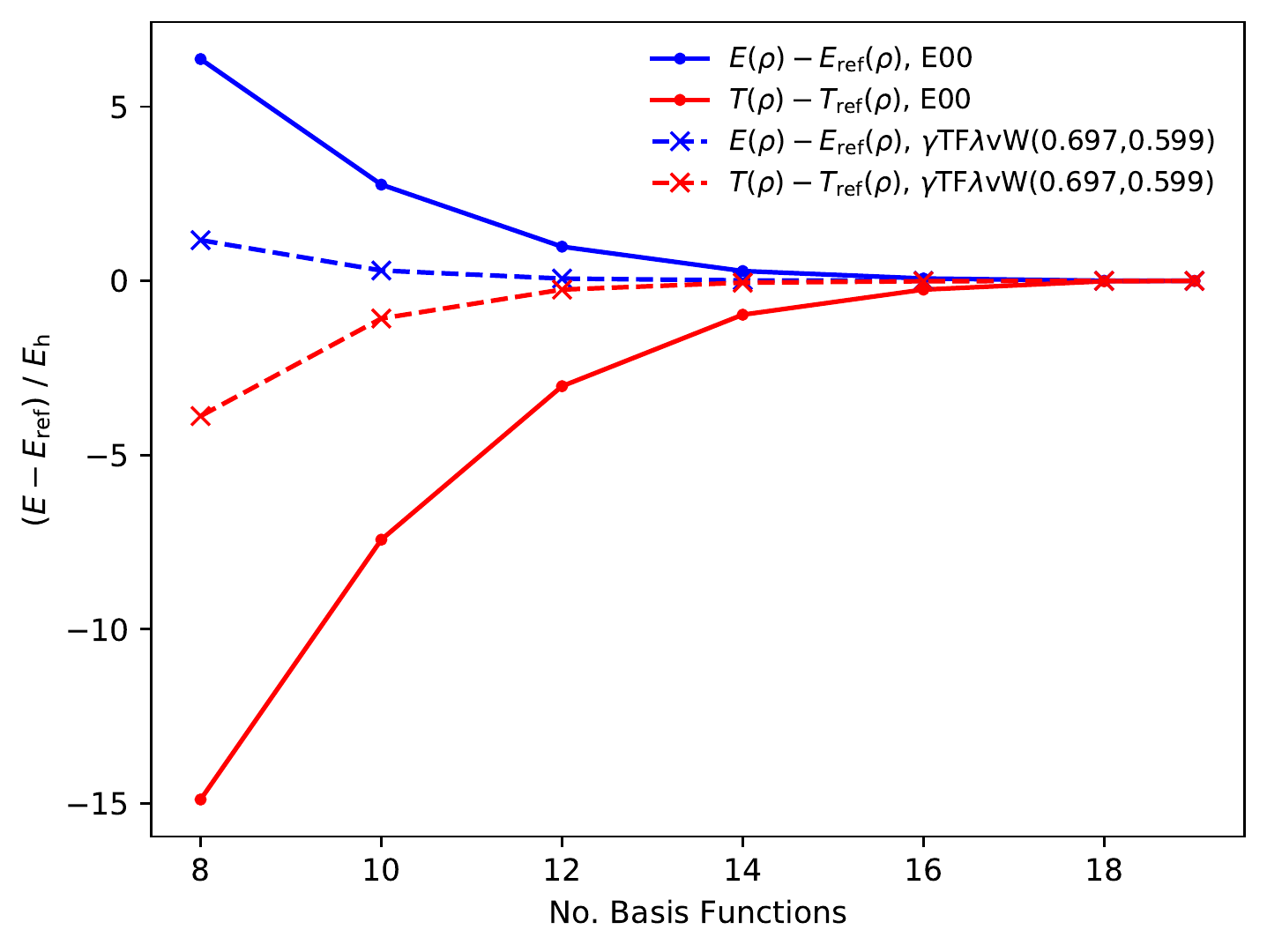}
\caption{Convergence of the energy (blue) and kinetic energy (red) for the neon atom in even tempered basis sets with between 8 and 19 s-type functions. The convergence of the $\gamma$TF$\lambda$vW functionals is in general more rapid than that of the GGA type functionals. The convergence for $\gamma$TF$\lambda$vW(0.697, 0.599) is shown by dashed lines, whilst the convergence for the GGA functional E00 is shown by solid lines. Convergence to better than 1\,m$E_\text{h}$ is achieved for 19 s-type functions.}\label{fig:bas_conv_atoms}
\end{figure}

The ability to converge rapidly in a given basis set is of little consequence if the convergence toward the basis-set limit cannot be achieved to a reasonable degree of accuracy. In Figure\,\ref{fig:bas_conv_atoms}, 
we show how the energy converges towards the basis-set limit for the $\gamma$TF$\lambda$vW(0.697, 0.599) functional. Convergence with respect to the basis size is relatively rapid, and we observe a similar pattern for other functionals in the $\gamma$TF$\lambda$vW family. Interestingly, the convergence for the GGA functionals is somewhat slower -- E00 is also shown in Figure\,\ref{fig:bas_conv_atoms}. 
For both the $\gamma$TF$\lambda$vW and GGA-type functionals, the inclusion of compact (high-exponent) functions is decisive in achieving convergence, with the GGA calculations requiring more compact functions to achieve tight convergence with respect to the basis set. As may be expected, the convergence properties of the $T(\rho)$ component of the energy mirror those of the total energy $E(\rho)$. For all the functionals considered here, the reference energy used to estimate the basis-set limit is obtained from a very large even tempered set of s-type functions with exponents defined as $1.6^n$ with $n = -20 \dots 35$. 
In each case, the 19s basis set used by CCH gives convergence relative to this limit of better than 1\,m$E_\text{h}$. 

\subsubsection*{Thomas--Fermi theory -- a pathological case}
Thomas--Fermi theory represents an important limiting case in DFT, becoming exact as $Z \rightarrow \infty$. It has been shown that the chemical potential $\mu$ in TF calculations is zero and that the energy of atomic systems is described by 
\begin{equation}
    E = -0.7687 Z^{7/3}. \label{ETF}
\end{equation}
The TF density varies with distance from the nucleus $r$ as $r^{-3/2}$ for $r \to 0$, diverging at the nucleus, and as $r^{-6}$ for $r \to \infty$. In addition, anions are unbound and, if applied to molecular systems, neither are molecules. A detailed exposition of TF theory can be found in Refs.\,\citenum{teller1962stability,Balazs,lieb1973thomas,lieb1977thomas,lieb1981thomas}. 

\begin{figure}
\includegraphics[width=\columnwidth]{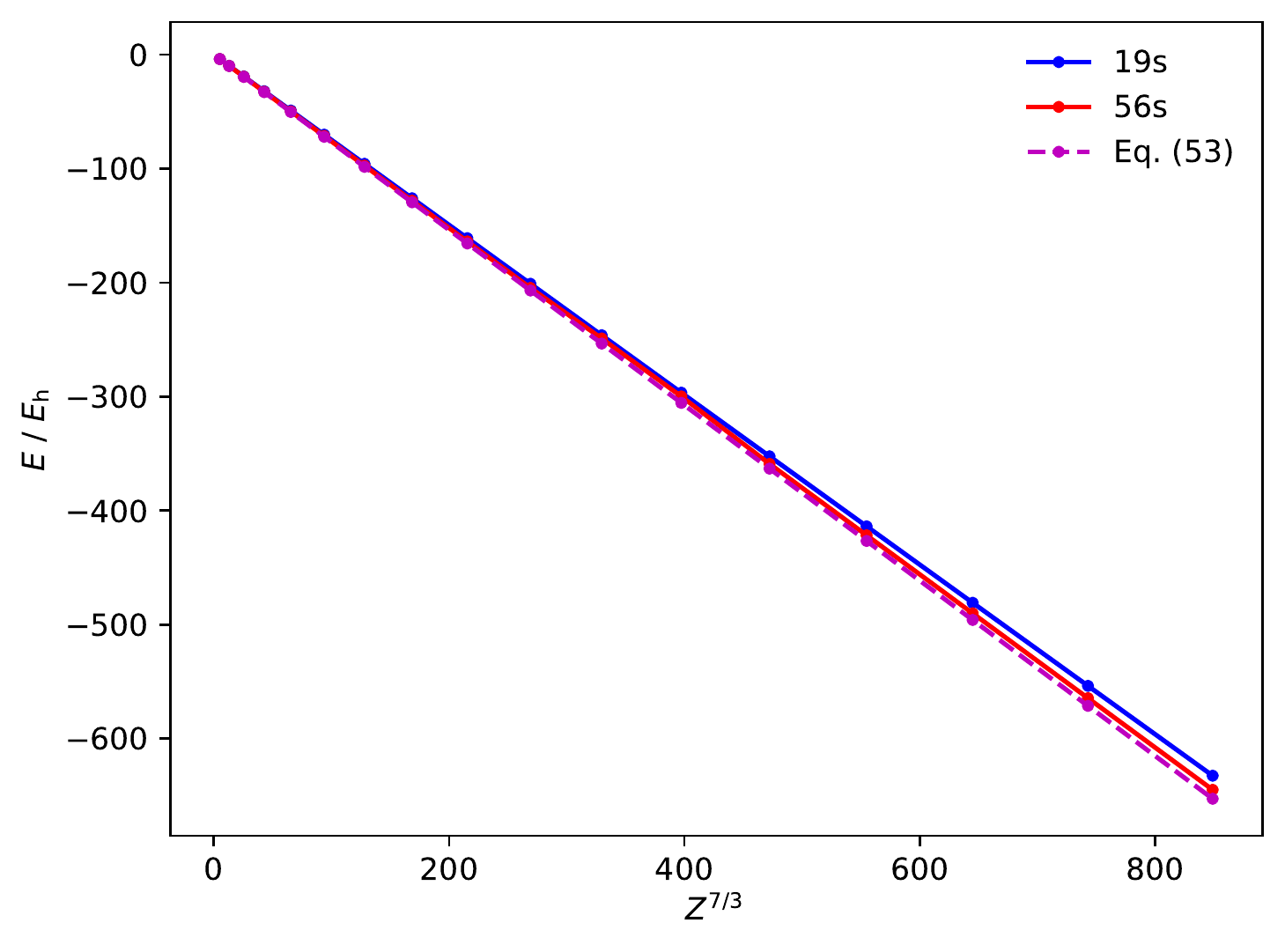}
\caption{The Thomas--Fermi energy calculated in even tempered finite basis with exponents $3.0^n$, $n = -6 \dots 12$ (blue) and $1.6^n$, $n = -20 \dots 35$ (red), compared with Eq.~(\ref{ETF}) (purple dashed line). Using the larger 56s basis set improves the energy for argon by more than $10$ $E_\text{h}$ compared to Eq.~(\ref{ETF}), however, the error remains at $7.8 E_\text{h}$.}\label{fig:tf}
\end{figure}

With such properties, calculations at the TF level present a pathological case for any finite-basis method. Nonetheless, as the basis set is increased, the energy should approach that in Eq.\,(\ref{ETF}), and it is interesting to explore to what extent this can be achieved. In Figure\,\ref{fig:tf}, we plot the energy as a function of $Z^{7/3}$ for atoms with $1 \leq Z \leq 18$. Qualitatively, both the 19s basis set used by CCH and an extensive basis of s-type Gaussians with exponents $1.6^n$ with $n = -20 \dots 35$ capture the scaling of the energy with $Z^{7/3}$. However, 
convergence toward the limit $-0.7687 Z^{7/3}$ of Eq.\,\eqref{ETF} is slow -- even in the largest basis set, a coefficient of $-0.7592$ is obtained, resulting in significant deviations from 
$-0.7687 Z^{7/3}$ for large $Z$. 
The basis-set convergence for the TF model for $T(\rho)$ is not improved by adding an exchange--correlation functional. We therefore do not consider the pure TF models further in this work. 

\subsection{Performance of kinetic-energy functionals for atoms}\label{subsec:atoms}
Atomic systems have been widely studied in OF-DFT previously, particularly for the $\gamma$TF$\lambda$vW family of functionals. Here, we have used the Dirac exchange functional in addition to the $\gamma$TF$\lambda$vW kinetic-energy functionals in order to facilitate comparison with previous studies. Furthermore, we consider five generalized gradient approximations: OL1, P92, E00, conjB86a and conjPW91. 

\subsubsection{TF--vW--Dirac-type functionals}
Results for atoms with $1 \leq Z \leq 18$ are presented in the supplementary information for $\gamma$TF$\lambda$vW with a range of ($\gamma$,$\lambda$) values. These include the vW functional~\cite{VWFUNC} ($0$,$1$), second-order gradient expansion~\cite{GE2a,GE2b} ($1$,$1/9$), Lieb's functional~\cite{lieb1981thomas} ($1$,$0.185909191$), the empirically optimized values~\cite{YT65} ($1$,$1/5$), Baltin's functional~\cite{Baltin} ($1$,$5/9$), the simple additive combination ($1$,$1$) and the optimized values of Lopez-Acevedo and co-workers~\cite{leal2015optimizing} ($0.697$,$0.599$). In each case, the resulting total energies and kinetic energies are presented. 

The approach widely used to assess the quality of kinetic-energy functionals is to evaluate them on a fixed density obtained from a Hartree--Fock or Kohn--Sham calculation. The value may then be compared against the usual Kohn--Sham $E(\rho)$ and $T_\text{s}(\{ \varphi_i \})$. Here, we use Dirac exchange-only Kohn--Sham calculations in the primitive def2-qzvp basis to generate values for $E(\rho)$ and $T_\text{s}(\{ \varphi_i \})$. 
For each functional, $T_\text{s}(\rho)$ and the corresponding $E(\rho)$ are then evaluated on the same density. 

The vW functional is a lower bound for the kinetic energy and the post-KS type analysis reflects this, with a systematic underestimation shown by a mean percentage error (MPE) of $-23.9$\% in the kinetic energy over this set of atoms, the underestimation increasing in magnitude with increasing $Z$. It should, of course, be kept in mind that the vW bound applies only when the functional is evaluated on the same density. 
Performing self-consistent calculations (using the even-tempered 19s basis set previously described), the MPE changes dramatically to 102.5\%. This overestimation does not represent a violation of the bound, but rather the fact that the self-consistent vW calculations become stationary at densities very far from the KS-DFT ones, typically being much more compact and having correspondingly high kinetic energies relative to the KS reference values. An exception to this trend is the helium atom for which, as a spin unpolarized two-electron case, the vW functional is exact. As would be expected from the basis-set convergence analysis, in this case the self-consistent solution agrees well (to better than 1\,m$E_\text{h}$) with the post-KS values, the difference simply reflecting the different choices of basis set in the calculations. 

\begin{figure*}
\includegraphics[width=0.5\textwidth]{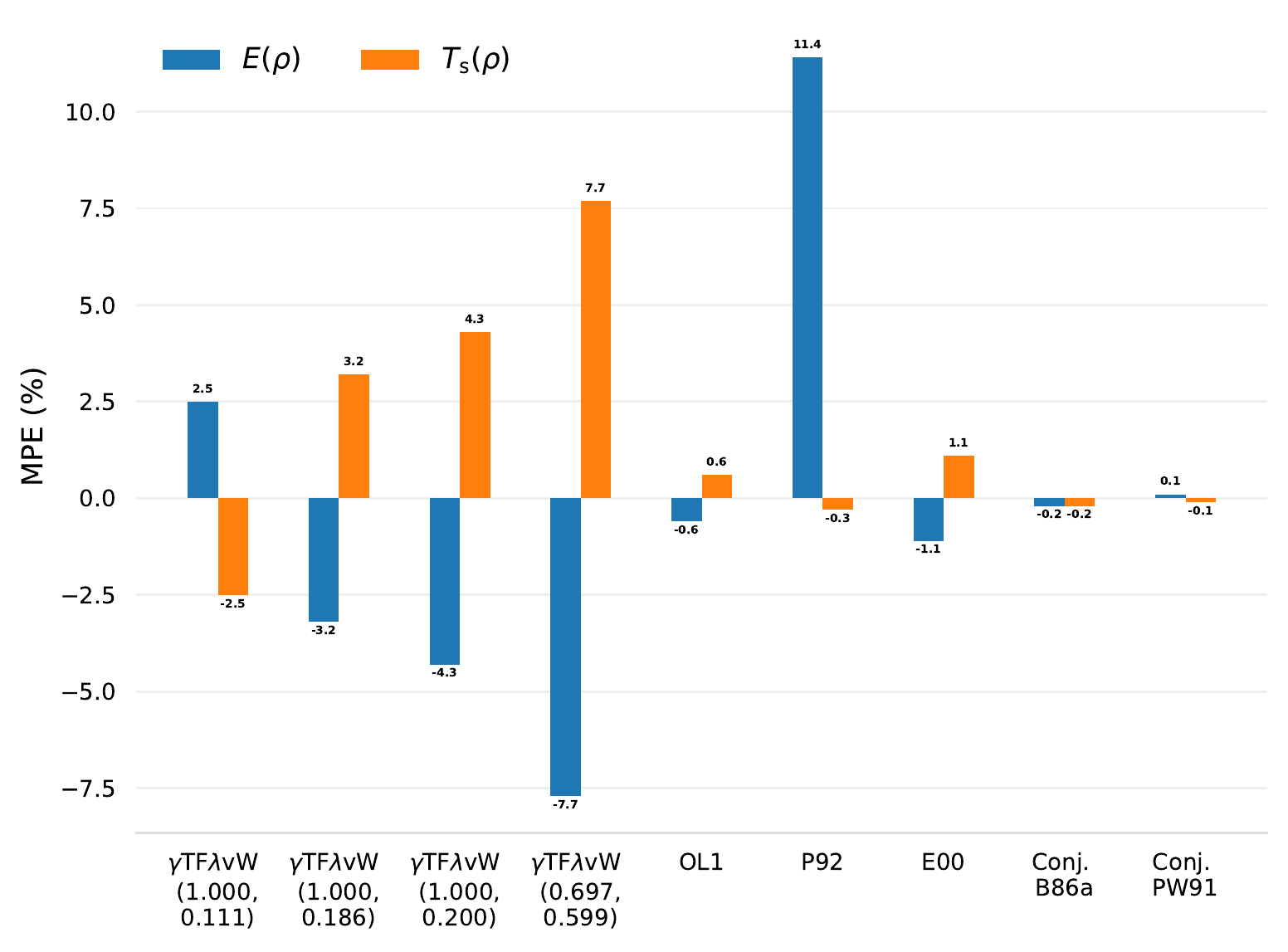}\includegraphics[width=0.5\textwidth]{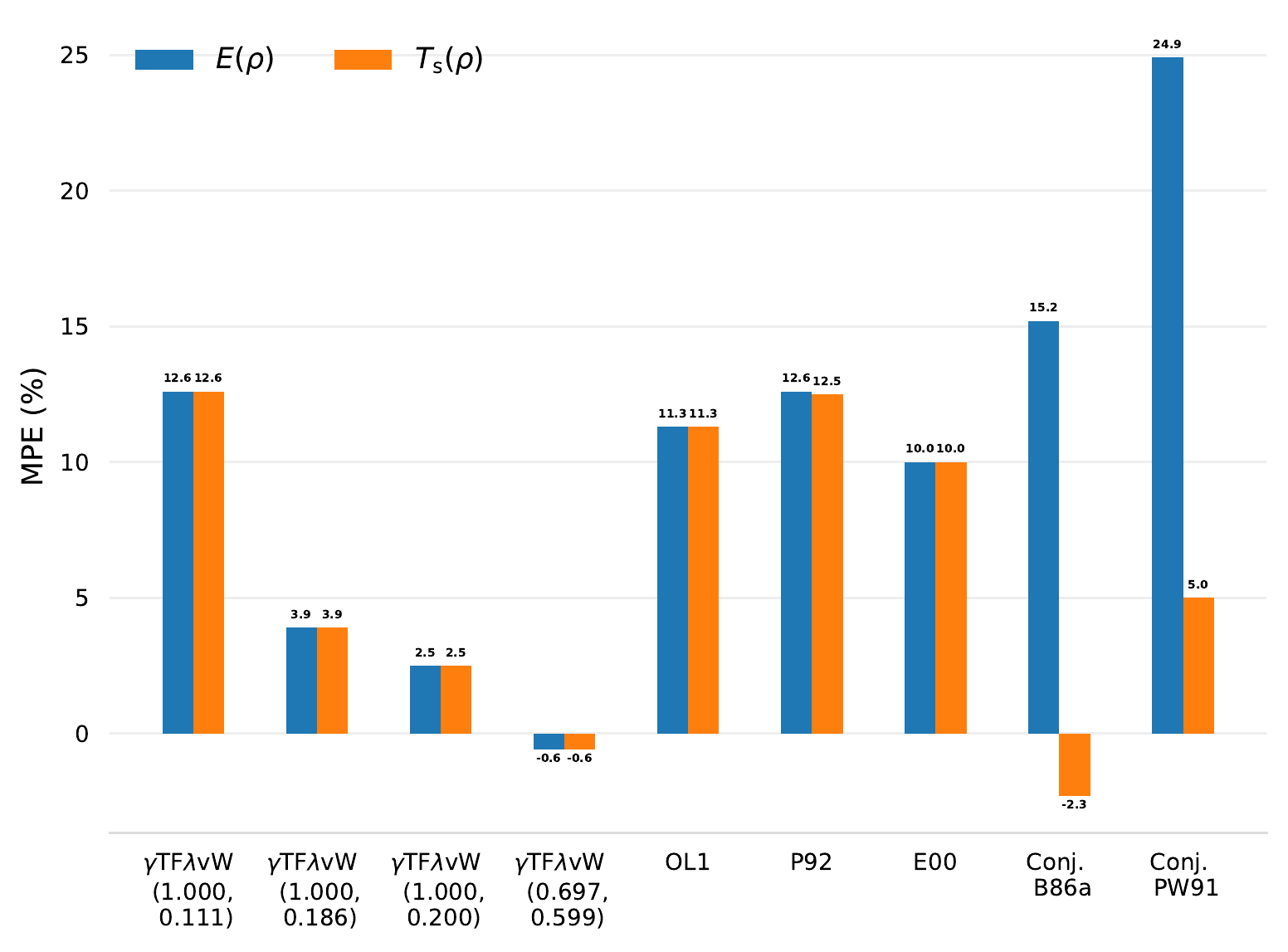}
\caption{Mean percentage errors in the energy (blue) and kinetic energy (orange) relative to Kohn--Sham LDA exchange-only calculations for atoms with $1 \leq Z \leq 18$. In the left panel, the analysis is carried on fixed densities obtained from Kohn--Sham LDA exchange-only calculations. In the right panel, the analysis is carried out on densities from self-consistent calculations. The difference is striking -- resulting in qualitatively different ranking of the functionals in terms of accuracy, with the $\gamma$TF$\lambda$vW functionals outperforming the GGA functionals in self-consistent evaluations, opposite to the conclusion that would be drawn based on fixed densities.}\label{fig:MPE_atoms}
\end{figure*}

The MPEs in $E(\rho)$ and $T_\text{s}(\rho)$ for the remaining $\gamma$TF$\lambda$vW functionals are presented in Figure~\ref{fig:MPE_atoms}. The vW functional is excluded due to its large MPE values, as are those with Baltin's value ($1$,$5/9$) and the simple additive combination ($1$,$1$), which give rise to post-KS MPEs of $31.4$\% and $65.2$\% respectively and self-consistent MPEs of $-19.8$\% and $-34.6$\% respectively. The left panel displays the post-KS error evaluation and the right panel the self-consistent error evaluation. 

In the post-KS analysis, the magnitude of the errors grows systematically as $\lambda$ increases. The optimized $\gamma$TF$\lambda$vW($0.697$,$0.599$) functional put forward by Lopez-Acevedo and co-workers~\cite{leal2015optimizing} shows errors in the total energy that are larger than the other $\gamma$TF$\lambda$vW functionals in Figure~\ref{fig:MPE_atoms}. However, the reduced $\gamma$ value gives significantly reduced errors compared to those obtained with, for example, with Baltin's values ($1$,$5/9$). It should also be noted that these values were chosen based on self-consistent calculations, not fixed densities.

In the right-hand panel of Figure~\ref{fig:MPE_atoms}, the MPEs from self-consistent calculations are presented. The difference is striking. The errors now systematically reduce as $\lambda$ increases when $\gamma=1$, up to $\lambda=1/5$, after which the errors rise again. Consistent with previous studies, the error in the total energy is minimized around $\lambda=1/5$ for these functionals. This  behaviour contrasts the post-KS measures, which would indicate $\lambda=1/9$ is optimal. If the value of $\gamma$ is also optimized in self-consistent calculations -- as done by Lopez-Acevedo and co-workers~\cite{leal2015optimizing} -- then the errors are further reduced to $-0.6$\% in the kinetic energy. This contrasts the error of $7.7$\% from the post-KS analysis. Together these results highlight the necessity of performing self-consistent calculations to meaningfully benchmark approximations to $T_\text{s}(\rho)$. Whilst screening of approximations for $E_\text{xc}(\rho)$ in a non-self-consistent manner is a useful tool to identify accurate models, see for example Ref.~\citenum{Cohen2012}, a similar practice is not justified for $T_\text{s}(\rho)$.
\begin{figure}
\includegraphics[width=0.5\textwidth]{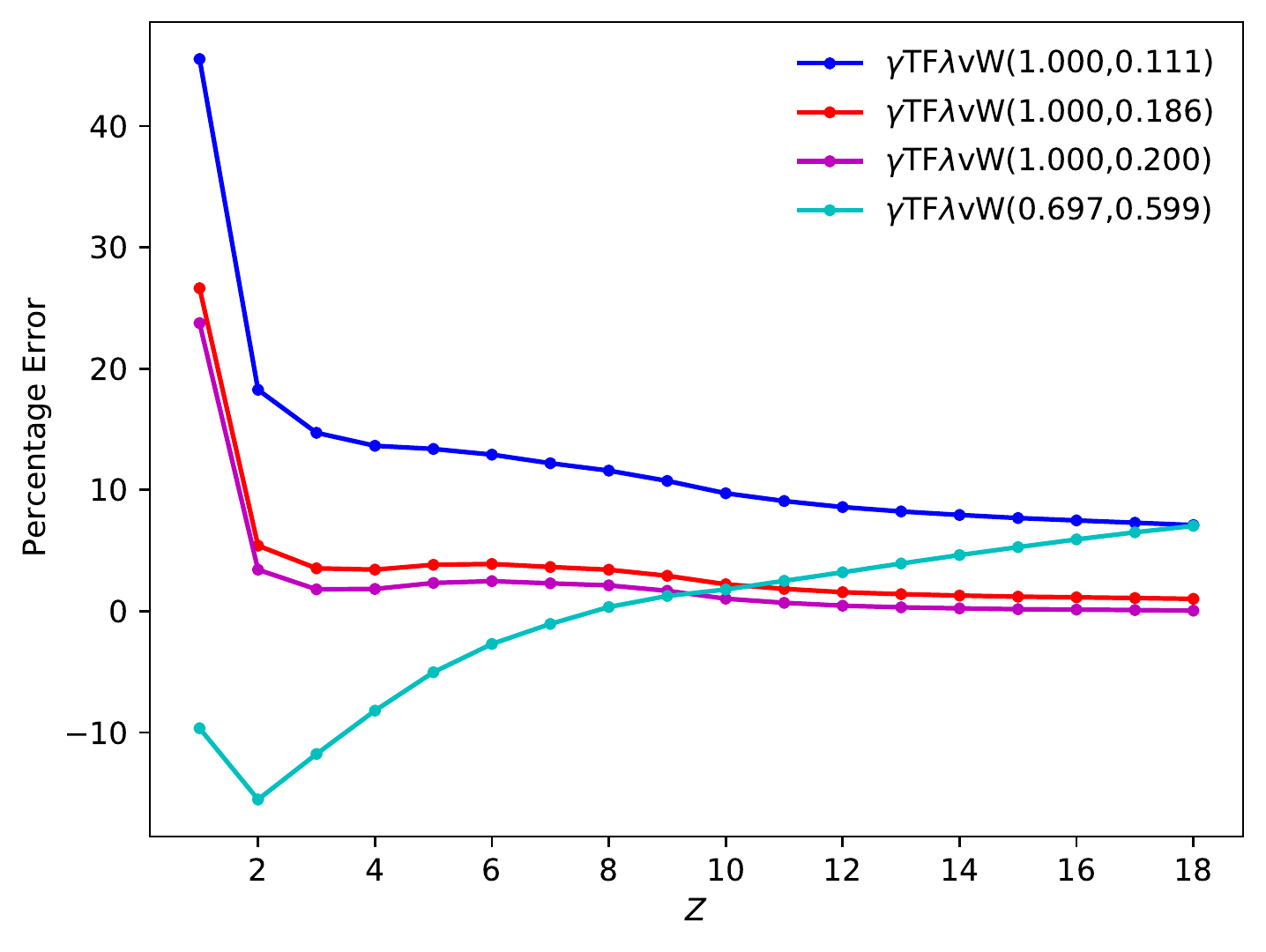}
\caption{Percentage errors for the atoms with $1 \leq Z \leq 18$ for the $\gamma$TF$\lambda$vW(1.000, 0.111), $\gamma$TF$\lambda$vW(1.000, 0.186), $\gamma$TF$\lambda$vW(1.000, 0.200) and $\gamma$TF$\lambda$vW(0.697, 0.599) functionals.}\label{fig:PE_atoms}
\end{figure} 

Whilst the MPEs give an average picture of the performance of each functional, it is important to consider if these are representative for all atoms. 
The percentage errors for each atom are shown in Figure~\ref{fig:PE_atoms} for a selection of the $\gamma$TF$\lambda$vW functionals. Larger percentage errors are obtained for $Z=1$, but in general the averages do reflect the relative quality of the functionals in each system, the exception being the optimized functional $\gamma$TF$\lambda$vW($0.697$,$0.599$). Whilst the MPEs are small over this data set, there is an error cancellation between atoms with $Z < 8$ and those with $Z > 8$. We have confirmed that, for higher $Z$, the percentage errors for this functional continue to increase and this error cancellation breaks down, perhaps reflecting the fact that this functional was fitted to atomic data with $1 \leq Z \leq 18$.

\subsubsection{Generalized gradient functionals}
The use of GGA-type models for $T_\text{s}(\rho)$ in a self-consistent manner has been discussed by Xia and Carter~\cite{Xia2015}, and by Trickey, Karasiev and Chakraborty~\cite{TrickeyCommXia2015}. Here we include just a few examples since, in line with the observations of these authors, many proposed GGAs are not stable when used self-consistently. 
For the atoms considered in this section, we are able to converge many functionals. However, for molecular systems, we make similar observations to Ref.\,\citenum{Xia2015}; many such functionals are not particularly stable, as such we include here only some of the most stable examples. 

The convergence issues for essentially all of these cases can be traced to poorly behaving $\delta T_\text{s}(\rho) / \delta \rho$, which for some functionals proposed in the literature are even divergent at some points in space.\cite{Borgoo2013,Borgoo2014} This observation again suggests that functional development and testing should be carried out using self-consistent evaluations, for which such issues become immediately clear. As noted by Trickey, Karasiev and Chakraborty~\cite{TrickeyCommXia2015}, the imposition of constraints in deriving the functional forms may alleviate these issues and result in functionals with greater numerical stability. In the present work, we find that our optimization schemes consistently converge functionals for which $\delta T_\text{s}(\rho) / \delta \rho$ is well behaved. 

The MPEs for the GGA functionals are shown in Figure~\ref{fig:MPE_atoms}. In the left panel, the post-KS error measures suggest that the OL1, E00, conjB86a and conjPW91 functionals all substantially improve over the $\gamma$TF$\lambda$vW functionals, with the conjoint functionals showing particularly good accuracy. Only the P92 functional has larger post-KS errors than the $\gamma$TF$\lambda$vW($1.000$, $0.111$) functional. The self-consistent MPEs are shown in the right panel. Again the trends are strikingly different, the magnitude of the errors for all of the GGAs significantly increases. For the OL1, P92 and E00 functionals, 
the MPEs for both the total energy and kinetic energy are above $10$\%. For the conjoint B86a and PW91 functionals, the total energies have greater than $10$\% MPE, but the kinetic energies 
have MPEs of $-2.3$\% and $5$\%, respectively. This observation again reinforces the need for self-consistent analysis. 

The convergence information for the GGAs is also presented in Table~\ref{tab:atom_conv}. Again, the mean, median, maximum and minimum number of steps required indicate rapid 
convergence, confirming the robustness of the TRIM optimization for atomic systems with a range of functional forms.

\subsubsection{Uniqueness of solutions}
In the discussion of the four-way correspondence in OF-DFT, the concave--convex nature of the saddle functional $\mathcal{E}(v,N)$ was highlighted. In particular, if the exact $\mathcal{L}_{v, N}(\rho, \mu)$ is used, 
then only one global first-order saddle point exists. Of course, when approximations to $\mathcal{F}(\rho)$ are employed, then it is interesting to investigate how many solutions can be obtained and whether the number of solutions reflects the quality of the approximations. To investigate this, we performed calculations on each atom with $1 \leq Z \leq 18$ using random initial guesses. For each atom, 25 random guesses were generated for the density expansion coefficients $c_i$, by assigning random values between 0 and 1 to each $c_i$ and then re-normalizing the vector of coefficients so that the initial guess gives a density that integrates to the target particle number $N$. Whilst this sampling is not exhaustive, it is sufficient to provide an indication of whether one, few or many solutions exist for a given functional.

In Table~\ref{tab:usol}, we show the number of solutions obtained for each atom for the functionals in this study, treating
solutions as distinct if the energies differ by more that $10^{-6}E_\text{h}$. All calculations are converged to $10^{-10}E_\text{h}$ or better;
the lowest-energy solutions and their chemical potentials are given in the supplementary information. TRIM was used for all the calculations presented in Table~\ref{tab:usol}, however, we have confirmed that the same solutions are obtained using the other optimization methods, given the same initial guess.

The results are quite revealing. For the $\gamma$TF$\lambda$vW functionals, very few solutions are found -- in many cases only one solution is obtained and in the worst cases no more than three solutions are obtained. 
Since the TRIM scheme uses the Hessian in the diagonal representation, all of the solutions are explicitly confirmed to be first-order saddle points. 
It is known that the use of TF and TFvW functionals gives approximations to $\mathcal{F}(\rho)$ are that are convex~\cite{lieb1977thomas, lieb1981thomas}. However, the addition of Dirac exchange leads to an approximation that is not convex. The observation that a number of solutions exist for $\gamma$TF$\lambda$vW functionals in combination with Dirac exchange is consistent with this. However, the low number of solutions found suggests that maintaining this property for approximations to $T_\text{s}(\rho)$, the dominant contribution to $\mathcal{F}(\rho)$, is advantageous from a numerical or practical point of view. It should be noted that a similar argument applies in Kohn--Sham theory, in which $T_\text{s}(\{ \varphi_i\})$ is constructed from the occupied Kohn--Sham orbitals $\{ \varphi_i\}$. Since $T_\text{s}(\{ \varphi_i\})$ is the exact form of $\mathcal{F}(\rho)$ in the noninteracting limit, it is convex by definition; this property is automatically incorporated into Kohn--Sham calculations. The introduction of approximate exchange--correlation functionals can result in the corresponding $\mathcal{F}(\rho)$ in Kohn--Sham calculations becoming nonconvex. However, since the largest component of this functional $T_\text{s}(\{ \varphi_i\})$ is necessarily convex, the nonuniqneness of solutions in Kohn--Sham DFT are expected to be less problematic than in OF-DFT.

The results for the GGA functionals are more varied. The OL1, P92 and E00 functionals behave in a similar manner to the $\gamma$TF$\lambda$vW family, with at most two solutions for a given atom. 
In contrast, the conjB86a and conjPW91 functionals give many solutions for every atom, demonstrating that these functionals are very far from concave--convex functionals. 
For this reason, we do not consider the conjB86a and conjPW91 functionals further in this work.  
\begin{table*}
\caption{The number of solutions obtained using 25 random sets of starting coefficients for each atom with $1 \leq Z \leq 18$. }\label{tab:usol}
\begin{tabular}{lrrrrrrrrrr}
\hline
\hline
Atom	 & & $\gamma$TF$\lambda$vW& $\gamma$TF$\lambda$vW & $\gamma$TF$\lambda$vW	&$\gamma$TF$\lambda$vW	&OL1	&P92	&E00	&conjB86a 	&conjPW91 \\
&$\gamma$   & $1.0$   & $1.0$  & $1.0$  & $0.697$  & & & & &  \\
&$\lambda$   &  $1/9$  & $0.186$ &  $1/5$ &  $0.599$ & & & & & \\
\hline
H	&&1	&2	&2	&2	&1	&1	&1	&22	&25\\
He	&&1	&1	&1	&2	&1	&1	&2	&24	&24\\
Li	&&1	&1	&1	&1	&2	&1	&2	&24	&24\\
Be	&&1	&1	&1	&1	&1	&1	&1	&18	&20\\
B	&&1	&1	&3	&1	&1	&1	&1	&17	&22\\
C	&&1	&3	&2	&1	&2	&1	&2	&14	&21\\
N	&&1	&2	&2	&1	&2	&1	&1	&15	&20\\
O	&&3	&2	&2	&1	&1	&1	&1	&14	&20\\
F	&&1	&2	&1	&1	&1	&2	&1	&12	&14\\
Ne	&&1	&1	&1	&1	&2	&1	&1	&13	&15\\
Na	&&1	&1	&1	&1	&2	&1	&1	&11	&12\\
Mg	&&1	&1	&1	&1	&1	&1	&1	&10	&10\\
Al	&&1	&1	&1	&1	&2	&2	&2	&13	&15\\
Si	&&1	&1	&1	&1	&2	&2	&1	&11	&13\\
P	&&1	&1	&2	&1	&1	&1	&1	&11	&15\\
S	&&1	&1	&1	&1	&2	&2	&1	&9	&11\\
Cl	&&1	&1	&2	&1	&1	&1	&1	&10	&10\\
Ar	&&1	&2	&1	&1	&1	&1	&1	&10	&7\\
\hline
Mean	&&1.1	&1.4	&1.4	&1.1	&1.4	&1.2	&1.2	&14.3	&16.6\\
\hline

\hline
\end{tabular}
\end{table*}

\subsection{Molecular systems}\label{subsec:mols}
In their study, CCH presented some results for diatomic molecules with the $\gamma$TF$\lambda$vW family of functionals. We have found that the CCH approach is quite robust and able to converge calculations for GGA functionals in addition to these. However, the nested optimization leads to many more steps than are required for the TRIM method. Here we consider a set of 28 small polyatomic molecules at the geometries given in 
Ref.\,\citenum{TealeNMR} and perform the calculations with the TRIM optimization scheme. 

\subsubsection{Superposition-of-atomic-densities initial guess}
For both the CCH and TRIM schemes, the rate of convergence is significantly improved if the optimization can be started close to the local region. To aid convergence, we form a superposition-of-atomic-densities (SAD) guess by calculating the density of each constituent atom in the same basis, and then forming a superposition of these for the initial guess. This approach leads to more rapid convergence in both the CCH and TRIM schemes. The atoms are calculated on the fly with the same functional, since each approximation leads to quite different (and often relatively unphysical) densities and so using the same functional for the SAD guess is necessary to obtain a good starting density. Furthermore, tabulating reasonable initial atomic densities for use in this setup is not feasible due to the widely different densities obtained with different approximations. 
For the chemical potential, we
have experimented with setting the initial value to different values derived from the atomic chemical potentials and have found that the mean of the atomic chemical potentials is a reasonable starting point. 

If the SAD guess is not used, the TRIM and CCH approaches still converge, but can spend many iterations in the global region before converging. For the TRIM approach, quadratic convergence is observed once the local region is reached. 

\begin{figure*}
\includegraphics[width=0.5\textwidth]{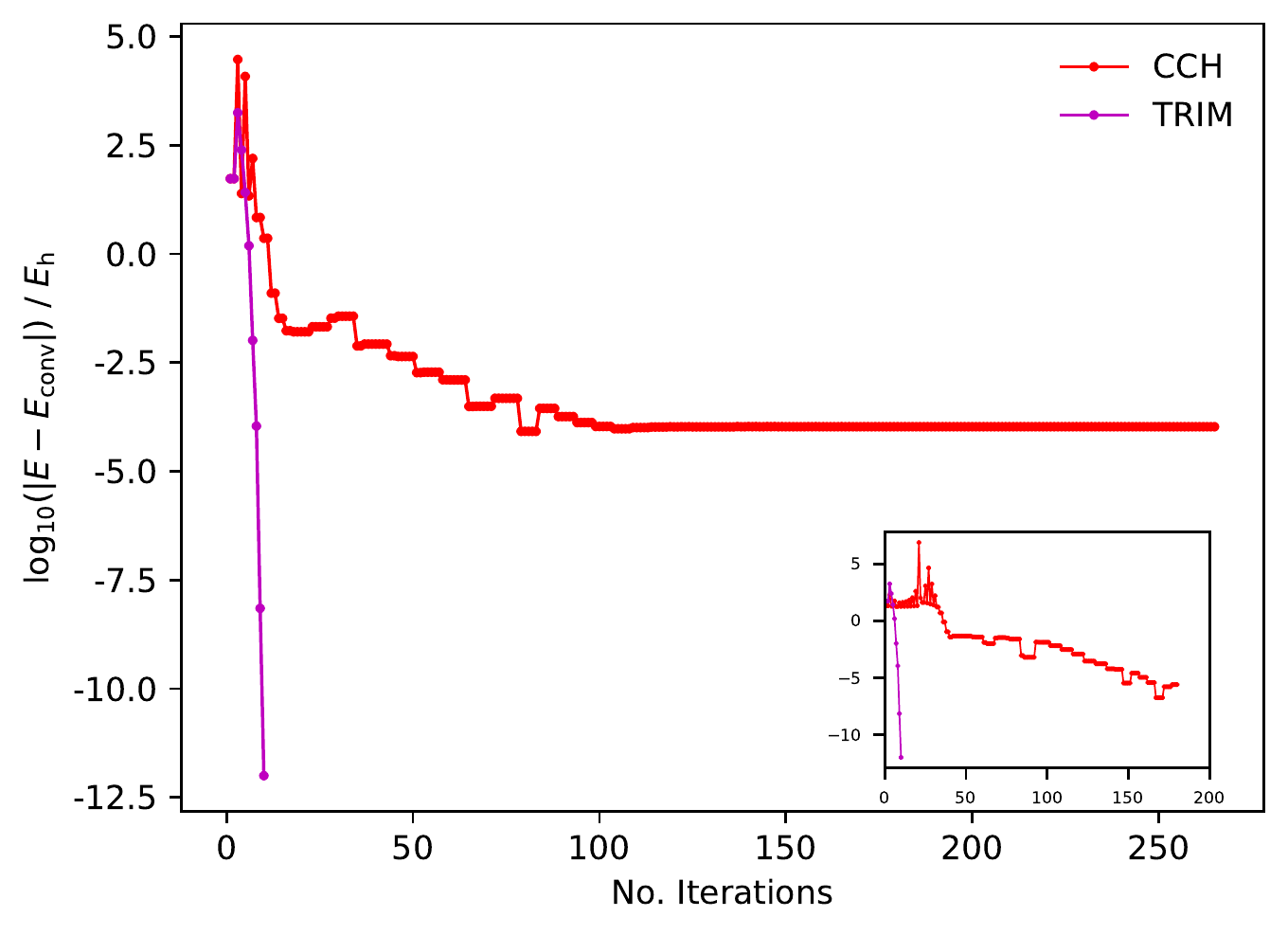}\includegraphics[width=0.5\textwidth]{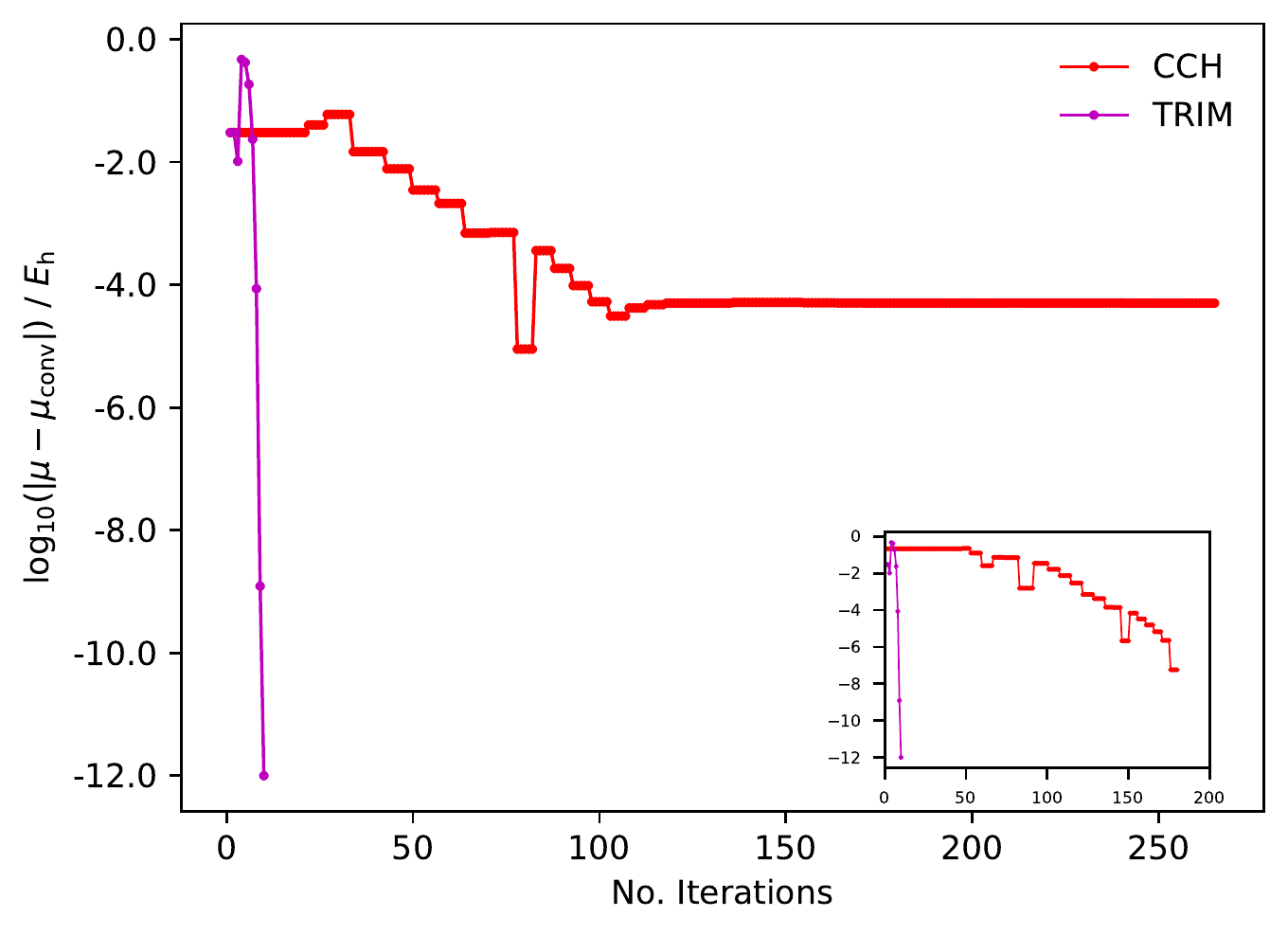}
\caption{Convergence of the CCH and TRIM optimization methods for the N$_2$ molecule, for the $(\gamma,\lambda) =$ ($0.697$,$0.599$) $\gamma$TF$\lambda$vW functional in the uncontracted def2-qzvp basis. In the left panel, the convergence of the energy is plotted as $\log_{10}(|E-E_\text{conv.}|)$ against the number of energy and gradient evaluations. In the right panel, 
the convergence of the chemical potential is plotted as  $\log_{10}(|\mu-\mu_\text{conv.}|)$ against the number of iterations. The insets show the same quantities for the nitrogen atom calculation with the same schemes, which is used to form the superposition of atomic densities guess.}\label{fig:n2_conv}
\end{figure*}

In Figure\,\ref{fig:n2_conv}, we show the convergence of the TRIM and CCH schemes for N$_2$ with the $\gamma$TF$\lambda$vW($0.697$,$0.599$) functional. In the left panel, 
the convergence of the energy is shown as a function of the iteration number; in the right panel, the convergence of the chemical potential is shown. 
In each case, the convergence for the preceding atomic calculation to generate the SAD guess is shown in the inset. 

With the conservative convergence criteria used here, the CCH scheme takes 180 iterations to converge the N atom and a further 265 iterations to converge 
the N$_2$ molecule. The reason for the slow convergence is the nested nature of the optimization and the number of bisection steps required to determine a well-converged value of the chemical potential. In contrast, the TRIM scheme takes just 8 iterations to converge the atomic calculation, followed by 11 steps to converge the molecular calculation. Moreover, the quadratic convergence of TRIM also affords much more precise convergence, with the error in the particle number reducing from $4.5 \times 10^{-4}$ to $2.8 \times 10^{-8}$ to $1.0 \times 10^{-12}$ in the last three steps, whereas the CCH scheme achieves a more modest particle number error of $1.3 \times 10^{-6}$. 

\subsubsection{Convergence of TRIM for molecular systems}
To gauge the transferability of the TRIM approach, we have performed calculations for 28 small molecules at the geometries in Ref.~\citenum{TealeNMR}, in uncontracted basis sets formed from the primitives of the def2-svp, def2-tzvp and def2-qzvp basis sets. In line with the observations for atoms, the basis-set convergence of the $\gamma$TF$\lambda$vW functionals is somewhat more rapid than for the OL1, P92 and E00 
GGA-type functionals. For example, we estimated the basis-set limit for the C$_2$H$_4$ molecule by performing calculations with a large even-tempered 19s6p5d3f2g basis set with exponents defined as $3^n$ 
with $-6\leq n \leq 12$ for s functions, $-3 \leq n \leq 2$ for p functions, $-2\leq n \leq 1$ for d functions, $-1\leq n \leq 1$ for f functions, and $-1\leq n \leq 0$ for g functions. For the $\gamma$TF$\lambda$vW($0.697$,$0.599$) and $\gamma$TF$\lambda$vW($1.000$,$0.200$) functionals, the uncontracted def2-qzvp basis deviates from this value by $0.5$ and $2.2$\,m$E_\text{h}$, respectively. For the OL1, P92 and E00 GGA functionals, 
the deviations are larger, falling between 8 and 10\,m$E_\text{h}$. In the present work, we regard the uncontracted def2-qzvp basis as a reasonable compromise between computational efficiency and  accuracy,
while allowing for comparison with Kohn--Sham LDA exchange-only calculations. Data for the largest basis set considered, u-def2-qzvp, are presented in the supplementary information.
\begin{table}
\caption{Convergence statistics for 28 small molecules at the geometries in Ref.~\citenum{TealeNMR}. The number of iterations corresponds to the number of evaluations of the Lagrangian, gradient and Hessian. The maximum, minimum, mean and median number of iterations to achieve convergence are presented for calculations in the uncontracted def2-svp, def2-tzvp and def2-qzvp basis sets.}
\begin{tabular}{lrrrrrr}
\hline
\hline
Basis	& &	$\gamma$TF$\lambda$vW &	$\gamma$TF$\lambda$vW &	OL1 &	P92 &	E00 \\
	&$\gamma$ &	$1.0$  &	$0.697$  &	 &	 &	 \\
	& $\lambda$ &	$1/5$ &	$0.599$ &	 &	 &	 \\
\hline
u-def2-svp	&Max &	47	&62	&79&	44	& 303 \\
	&Min	 &9	& 9	& 9	& 9	& 11 \\
	&Mean &	18	& 17 & 	22	& 18	&46 \\
	&Median &	16	&14	&17	&16	&25 \\
u-def2-tzvp	&Max	& 106	&36	&153	 & 69 	&113\\
	&Min	&12	&11	&12	&12	&13 \\
	&Mean	&26	&17	&30	&21	&31\\
	&Median	&22	&15	&18	&17	&24 \\
u-def2-qzvp	&Max	&875	 &629 	&962	 &373&	744 \\
	&Min	&12	&12	&14	&13	&14 \\
	&Mean	&55	&45	&102 &	76	&93 \\
	&Median	&19	&15	&23	&26	&24 \\
\hline
\hline
\end{tabular}
\end{table}

The convergence properties of GGA functionals have been discussed previously by Xia and Carter~\cite{Xia2015} and by Trickey, Karasiev and Chakraborty~\cite{TrickeyCommXia2015}. 
Whilst for atoms most reasonable GGA functionals can be converged easily, the situation is more challenging for molecules. Here we consider the $\gamma$TF$\lambda$vW functionals with $(\gamma,\lambda) =$ ($1.0$,$0.2$) and ($0.697$,$0.599$) and the GGA functionals OL1, P92 and E00. For the 28 small molecules considered, we find that convergence can be achieved for all of these functionals.

\subsubsection{$\gamma$TF$\lambda$vW and GGA accuracy for molecules}

\begin{figure}
\includegraphics[width=\columnwidth]{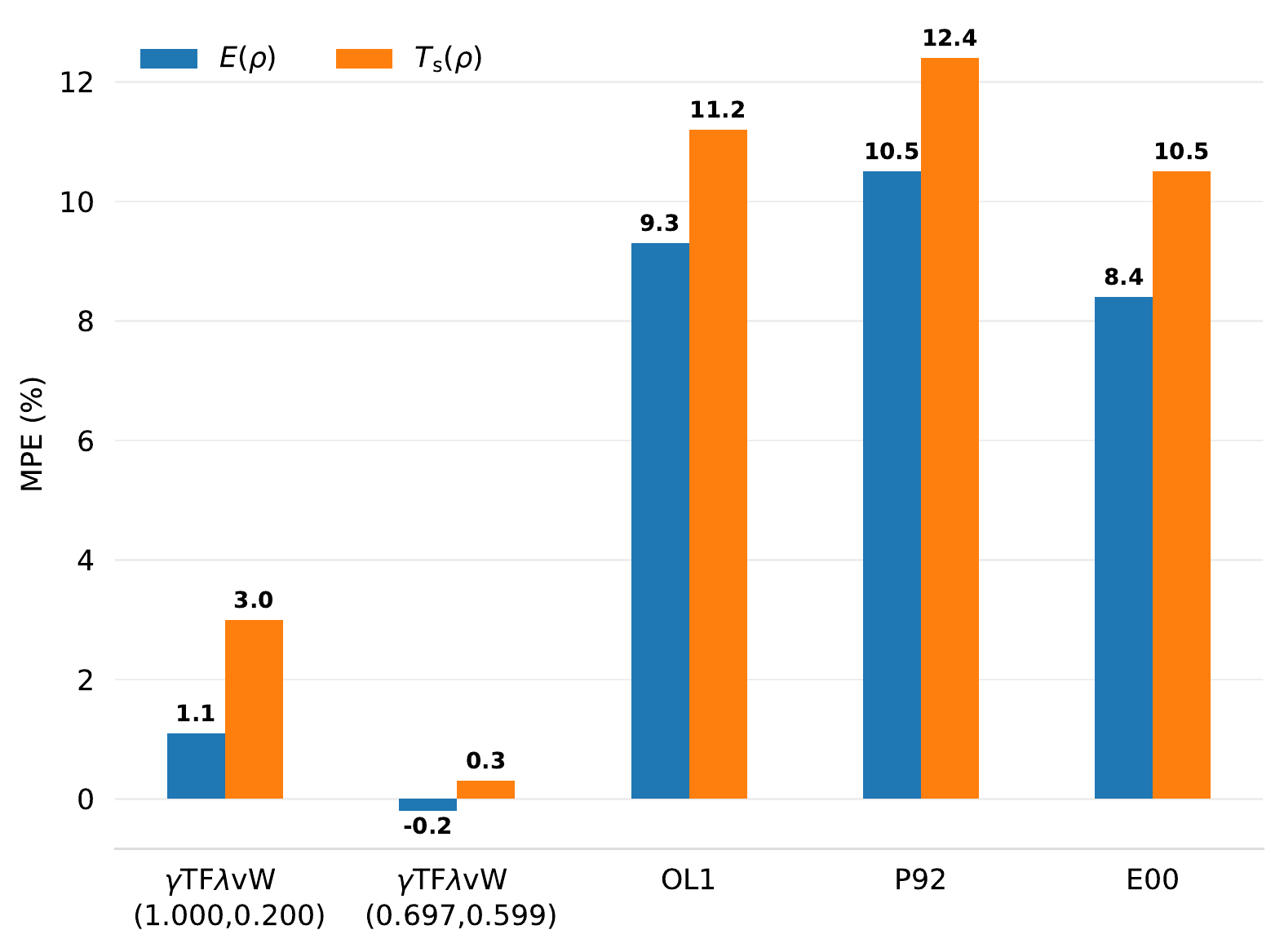}
\caption{Mean percentage errors in the total energies (blue) and kinetic energies (orange) for the set of 28 small molecules considered in the is work, relative to Kohn--Sham LDA exchange-only calculations. The errors for $(\gamma,\lambda) =$ ($0.697$, $0.599$) $\gamma$TF$\lambda$vW are lowest consistent with the results for atomic systems}\label{fig:MPEmol}
\end{figure}

In general, the trends for molecules parallel those for atoms; see Figure~\ref{fig:MPEmol} for the MPEs of $\gamma$TF$\lambda$vW with $(\gamma,\lambda) =$ ($1.000$,$0.200$) and ($0.697$,$0.599$), along with the 
OL1 and E00 GGA functionals. For the $\gamma$TF$\lambda$vW family, the $(\gamma,\lambda) =$ ($1.000$, $0.200$) functional has low MPEs for $E(\rho)$ and $T_\text{s}(\rho)$ of $-2.9$\% and $-7.0$\%, respectively, relative to KS calculations with Dirac exchange in the same basis. 
A comparison of the optimized $\mu$ value with $-\varepsilon_\text{HOMO}$ highlights the large errors in the chemical potential, with the OF-DFT functionals tending to strongly underestimate $\mu$, yielding an MPE of $-73.0$\%. 

The $\gamma$TF$\lambda$vW functional
$(\gamma,\lambda) =$ ($0.697$,$0.599$) was optimized both for energies and against the KS HOMO for atoms. The molecular errors for $E(\rho)$ and $T_\text{s}(\rho)$ for this functional reduce to $-1.5$\% and $-1.4$\%, respectively. More significantly, the MPE for $\mu$ is reduced to $-17.1$\% which, whilst still large, is significantly smaller than any of the other functionals considered here. 

For the OL1 and E00 functionals, similar performances are obtained. For the OL1 functional, the MPEs for $E(\rho)$ and $T_\text{s}(\rho)$ are $2.4$\% and $-3.9$\%, respectively, whilst for $\mu$ the MPE is $-75.4$\%; for the E00 functional, the MPEs for $E(\rho)$ and $T_\text{s}(\rho)$ are $2.3$\% and $-4.0$\%, respectively, whilst for $\mu$ the MPE is $-65.0$\%. It is clear that, for molecular systems, the $\gamma$TF$\lambda$vW($0.697$, $0.599$) functional is the most accurate. However, the errors relative to the energies of KS with Dirac exchange suggest that the energies are still not sufficiently accurate for chemical applications.   
  
\subsubsection{Molecular binding and size-consistency}
\begin{figure*}
\includegraphics[width=0.5\textwidth]{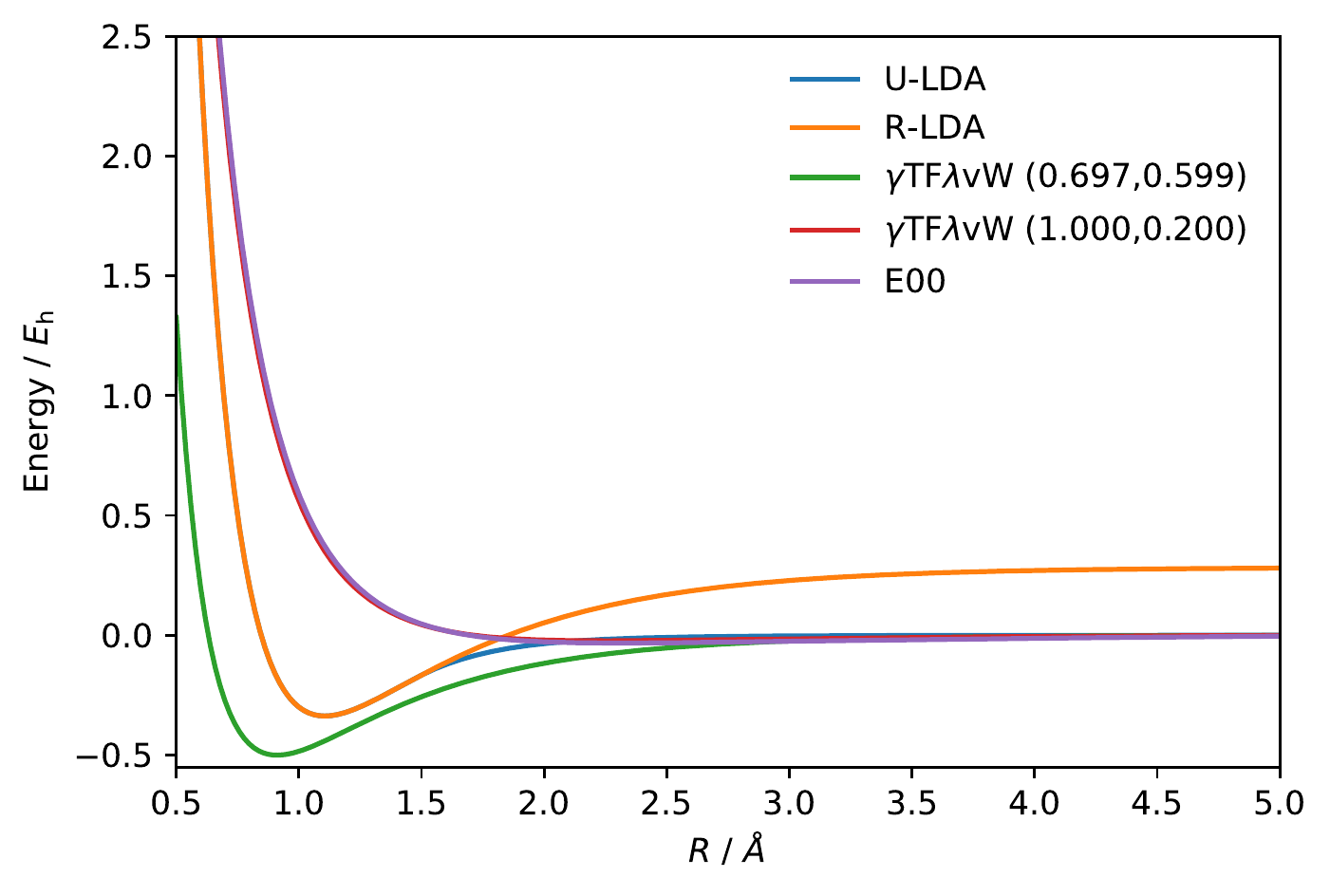}\includegraphics[width=0.5\textwidth]{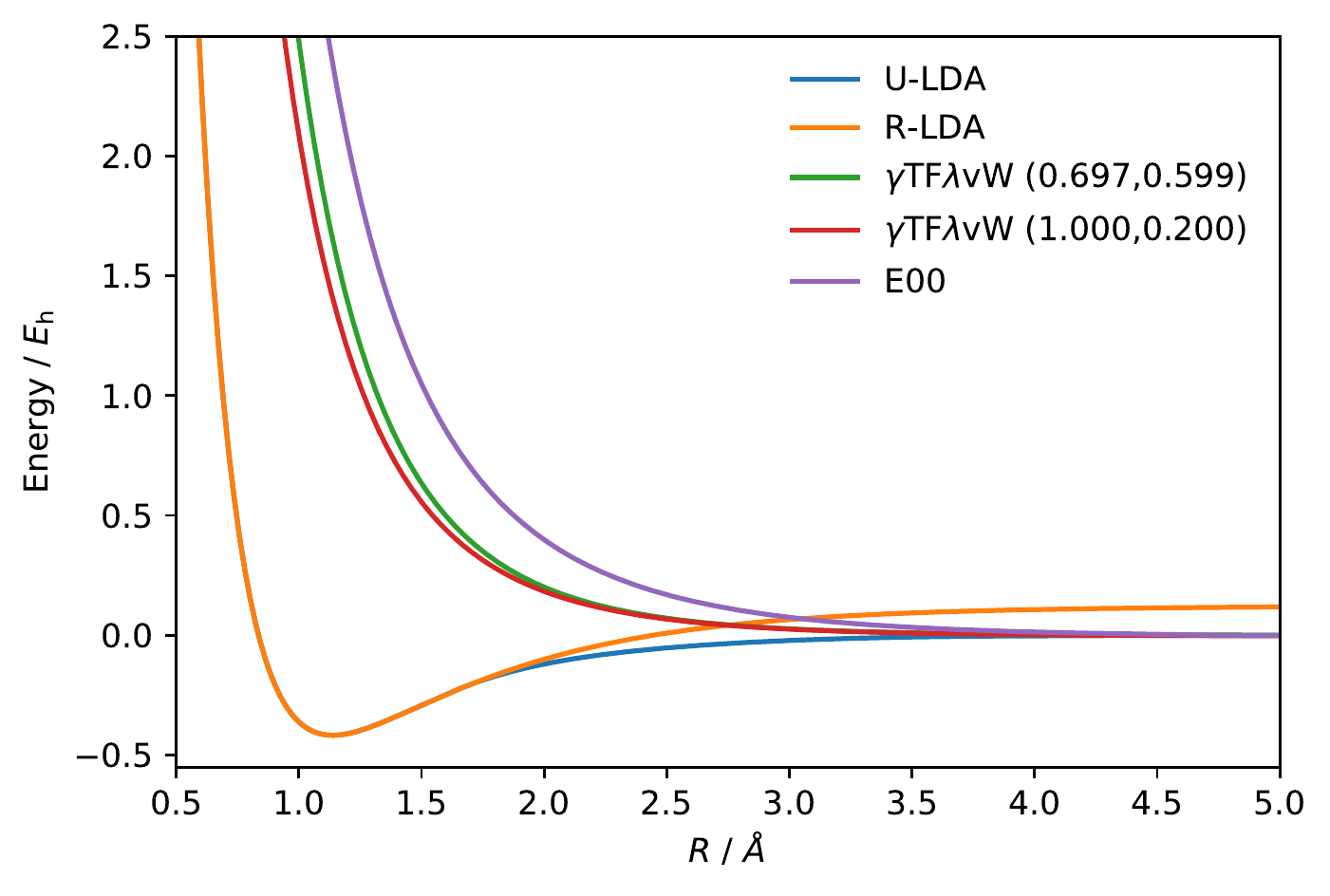}
\caption{The binding energy as a function of internuclear separation, $R$ / \AA, for N$_2$ (left panel) and CO (right panel).}\label{fig:PES}
\end{figure*}

It is well known that Thomas--Fermi theory does not lead to molecular binding. Unfortunately, as shown in Figure\,\ref{fig:PES}, for the examples of N$_2$ and CO, 
the situation is not improved by GGA functionals. Interestingly, the $\gamma$TF$\lambda$vW($0.697$,$0.599$) functional does yield some binding in many cases. 
However, this behaviour is far from consistent; for example, the N$_2$ molecule is bound with $R_\text{eq} = 0.915$\,\AA, but the isoelectronic CO has a repulsive potential energy surface. 
We have implemented analytic gradients for functionals up to the Laplacian level in \textsc{Quest}, however, for the functionals considered here (with the possible exception of $\gamma$TF$\lambda$vW($0.697$,$0.599$)), 
this capability is of limited utility. Nonetheless, some GGA functionals have been developed that favour more accurate binding energies at the expense of accuracy of the absolute energies~\cite{karasiev2006born}. 
The availability of analytic gradients may therefore be useful for functional development in future work. 

As noted by CCH, the spin-unpolarized formulation of OF-DFT may not be size-consistent, much like the restricted Kohn--Sham formalism. In Figure~\ref{fig:PES}, the OF-DFT curves approach zero for N$_2$, 
since this homonuclear diatomic molecule dissociates into two atoms with equal chemical potential. However, for CO, the curves approach a value slightly above zero, indicating violation of 
size-consistency by about 2\,m$E_\text{h}$. The average of the chemical potentials for the isolated C and O atoms is $-0.0433E_\text{h}$, whilst the chemical potential for the CO molecule at 500\,bohr separation 
is $-0.0474E_\text{h}$. A spin-polarized formulation of OF-DFT may restore size-consistency. This would require searches for higher-order saddle points, 
to which the TRIM formulation may be readily extended. Extensions to determine higher-order saddle points in the the context of geometry optimization have been reported recently.\cite{ALTRUISM}

\section{Conclusions}\label{sec:conc}
We have presented a second-order optimization approach based on the trust-region image method (TRIM). 
This approach is robust and offers more rapid convergence than comparable methods when implemented in an all-electron Gaussian basis setting. In particular, it offers significantly faster convergence 
than the previously presented  nested optimization by Chan, Cohen and Handy~\cite{Chan2001}. The second-order nature of this approach ensures that the solutions can be confirmed as first-order saddle-points, 
with quadratic convergence in the local region. The price to be paid for the improved convergence of TRIM is the diagonalization of the Hessian.
Future work will consider alternative representations of the Hessian, thereby restoring the intrinsic linear scaling of the OF-DFT approach. Our present implementation is in the context of all-electron optimization using Gaussian functions for expansion of the density. However, the optimization algorithm is not specific to problems of this kind and could feasibly be applied in other contexts such as real-space or plane-wave based OF-DFT implementations.

In the present work, the TRIM method shows that simultaneous optimization of the chemical potential and the energy can be an efficient and robust approach to the solution of the Euler--Lagrange equation central to OF-DFT. This approach allows us to investigate the stability of proposed energy functionals with some confidence, which has been a point of discussion in the literature. 
Our findings in this regard are in line with the conclusions of Xia and Carter,\cite{Xia2015} underlining the necessity for the evaluation of OF-DFT functionals in a self-consistent manner. 
In particular, the energy functionals must have well-defined functional derivatives and 
the typical pre-screening of functionals in a non-self-consistent manner, as is common for exchange--correlation functionals, is not appropriate for kinetic-energy functionals.

The four-way correspondence and the concave--convex saddle-point nature of the objective functional pertaining to OF-DFT shows that an accurate functional should have one global stationary point. Using the TRIM method, we have performed many optimizations starting from random guesses. Interestingly, functionals based on the conjointness conjecture~\cite{Lee1991} show many solutions, whereas $\gamma$TF$\lambda$vW functionals tend to yield only a few solutions. Future work will examine the design of functionals for which the density component of the Hessian is positive definite by construction. 
For GGA functionals, Trickey and co-workers~\cite{TrickeyCommXia2015} have already suggested constraints that may improve stability in practical optimization. Combination with constraints from density-scaling relations may also prove useful in deriving useful forms~\cite{Borgoo2012,Borgoo2013,Borgoo2014,Borgoo2014a}.
In the present work, we have highlighted how the TF limit is difficult to reach accurately in finite-basis methods, thus suggesting that GGA enhancement factors should be designed to ensure that this limit is carefully managed for typical molecular densities.

The empirically optimized $\gamma$TF$\lambda$vW($0.697$,$0.599$) functional of Lopez-Acevedo and co-workers~\cite{leal2015optimizing} is the most accurate studied in the present work. 
In general, we have not been able to find a numerically stable GGA functional that outperforms this simple functional. However, the platform presented here, combined with a robust optimization 
procedure and the simple implementation of functionals using the automatic differentiation techniques in the XCFun package~\cite{XCFun} provides a useful basis for further exploration. 
In combination with the observations on the concave--convex nature of the objective function in this work, further investigation of GGA-type corrections may be worthwhile. 
More recently, a range of Laplacian-based functionals have been proposed~\cite{laricchia2013laplacian,smiga2017laplacian}. We have implemented the required functional derivatives for the gradient and Hessian of Laplacian dependent functionals and work on their self-consistent evaluation is underway. Given that numerical difficulties with Laplacian--based functionals have been observed with exchange--correlation functionals~\cite{Neumann1997,Knowles1995}, we expect that a robust optimization scheme such as TRIM will be very useful in this context.

It is clear that even the most accurate approaches in the present work are not sufficiently accurate for general chemical application for finite molecular systems in an all-electron treatment. 
In essence, at least an order of magnitude improvement in the accuracy of $T(\rho)$ models will be required for OF-DFT to become chemically useful. 
In the context of solid-state applications to crystalline materials, much progress has been made using nonlocal kinetic-energy functionals. 
Many of these methods are formulated naturally in the solid-state context, but we note that procedures have been suggested for translation to real-space implementations~\cite{Choly2002}. With the development platform used in the present work, nonlocality may be explored directly in real space; work in this direction is also underway. 

Finally, we note that the nature of the Euler--Lagrange equation suggests the importance of the potentials (functional derivatives with respect to the electron density) in determining convergence. 
The significance of the kinetic potential was highlighted by King and Handy~\cite{king2000kinetic} along with a scheme for extracting this from standard Kohn--Sham calculations 
in all-electron Gaussian basis sets. Furthermore, techniques for the optimization of the Lieb functional~\cite{Lieb1983}, which have also been established as practical tools in this 
context~\cite{Savin1999,Yang2003,Teale2009,Teale2010}, provide access to the potentials entering the Euler equation for highly accurate densities. 
We expect that the TRIM approach for approximate functionals, presented in this work, in combination with these techniques will provide a useful tool in the development and testing of new OF-DFT approaches. 

\section*{Supplementary Information}
Energies for the atomic and molecular systems with the kinetic-energy functionals considered in this work are tabulated in the supplementary information. 

\section*{Acknowledgements}
We acknowledge financial support from the European Research Council under H2020/ERC Consolidator Grant top DFT (Grant No.\ 772259). We also acknowledge support from the Engineering and Physical Sciences Research Council (EPSRC), Grant No.\ EP/M029131/1. This work was partially supported by the Research Council of Norway through its Centres of Excellence scheme, project number 262695. We are grateful for access to the University of Nottingham's Augusta HPC service.  

\bibliography{paper}

\newpage
\section*{TOC Graphic}
\begin{figure}
\includegraphics[width=9.3cm]{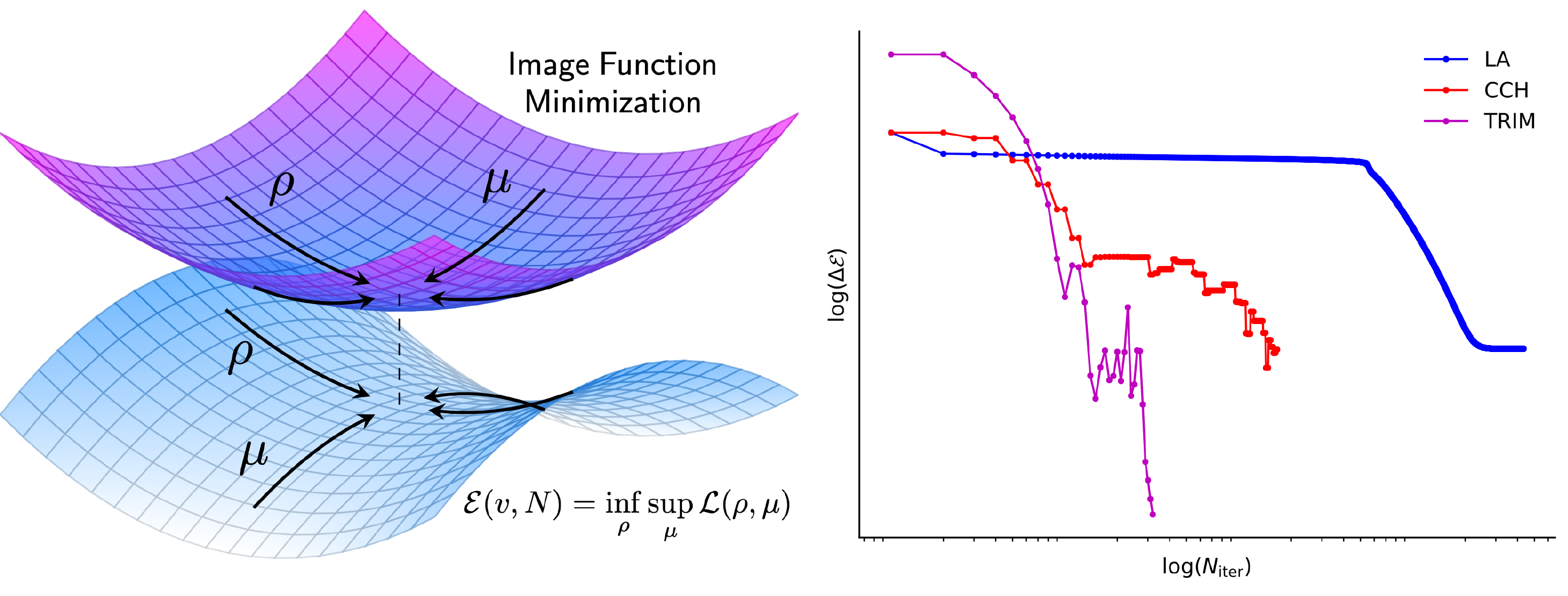}
\end{figure}

\end{document}


\date{\today}
\maketitle
\section{Atomic Data}
All calculations were carried out in a 19s even-tempered basis set with exponents $3^n$, $n = -6 \dots 12$. In each case 25 random starting guesses were generated and the lowest energy solution is presented in the table. The number of uniques solutions (to a tolerance of $10^{-6} E_\text{h}$) is also given, along with the number of TRIM iterations required for convergence. 
\begin{table}[h]
\caption{Energies, kinetic energies, chemical potentials and percentage errors relative to Kohn-Sham LDA exchange-only calculations using the $\gamma$TF$\lambda$vW(1.0, 1/9) functional. The number of uniques solutions generated from 25 random starting guesses is given, along with the number of iterations required for convergence.}
\begin{tabular}{lrrrrrrr}
\hline
\hline
Atom & E / $E_\text{h}$   & E \% Err & KE / $E_\text{h}$       & KE \% Err & $\mu$ / $E_\text{h}$        & $N_\text{sol}$ & $N_\text{iter}$ \\
\hline
H    & -0.6664   & 45.8     & 0.6663   & 45.5      & -0.059154 & 1      & 9     \\
He   & -3.2228   & 18.3     & 3.2227   & 18.3      & -0.061522 & 1      & 10    \\
Li   & -8.2515   & 14.7     & 8.2515   & 14.7      & -0.062519 & 1      & 10    \\
Be   & -16.1631  & 13.6     & 16.1632  & 13.6      & -0.063092 & 1      & 10    \\
B    & -27.2876  & 13.4     & 27.2877  & 13.4      & -0.063491 & 1      & 11    \\
C    & -41.9051  & 12.9     & 41.9053  & 12.9      & -0.063777 & 1      & 31    \\
N    & -60.2621  & 12.2     & 60.2622  & 12.2      & -0.063999 & 1      & 13    \\
O    & -82.5794  & 11.6     & 82.5792  & 11.6      & -0.064200 & 3      & 12    \\
F    & -109.0587 & 10.7     & 109.0577 & 10.7      & -0.064365 & 1      & 11    \\
Ne   & -139.8855 & 9.7      & 139.8835 & 9.7       & -0.064523 & 1      & 22    \\
Na   & -175.2328 & 9.1      & 175.2290 & 9.1       & -0.064663 & 1      & 20    \\
Mg   & -215.2623 & 8.6      & 215.2561 & 8.6       & -0.064787 & 1      & 32    \\
Al   & -260.1268 & 8.2      & 260.1171 & 8.2       & -0.064915 & 1      & 12    \\
Si   & -309.9706 & 7.9      & 309.9562 & 7.9       & -0.065021 & 1      & 30    \\
P    & -364.9313 & 7.7      & 364.9105 & 7.7       & -0.065128 & 1      & 39    \\
S    & -425.1398 & 7.5      & 425.1108 & 7.5       & -0.065230 & 1      & 35    \\
Cl   & -490.7217 & 7.3      & 490.6822 & 7.3       & -0.065328 & 1      & 13    \\
Ar   & -561.7975 & 7.1      & 561.7448 & 7.1       & -0.065421 & 1      & 13    \\
\hline
Mean     &           & 12.6     &          & 12.6      &           &        & 19   \\
\hline
\hline     
\end{tabular}
\end{table}

\begin{table}[h]
\caption{Energies, kinetic energies, chemical potentials and percentage errors relative to Kohn-Sham LDA exchange-only calculations using the $\gamma$TF$\lambda$vW(1.0, 0.186) functional. The number of uniques solutions generated from 25 random starting guesses is given, along with the number of iterations required for convergence.}
\begin{tabular}{lrrrrrrr}
\hline
\hline
Atom & E / $E_\text{h}$   & E \% Err & KE / $E_\text{h}$       & KE \% Err & $\mu$ / $E_\text{h}$        & $N_\text{sol}$ & $N_\text{iter}$ \\
\hline
H    & -0.5798   & 26.9     & 0.5798   & 26.6      & -0.064686 & 2      & 9     \\
He   & -2.8724   & 5.5      & 2.8724   & 5.4       & -0.068976 & 1      & 10    \\
Li   & -7.4476   & 3.5      & 7.4476   & 3.5       & -0.070841 & 1      & 11    \\
Be   & -14.7108  & 3.4      & 14.7109  & 3.4       & -0.071970 & 1      & 13    \\
B    & -24.9886  & 3.8      & 24.9888  & 3.8       & -0.072755 & 1      & 30    \\
C    & -38.5591  & 3.9      & 38.5593  & 3.9       & -0.073370 & 3      & 20    \\
N    & -55.6667  & 3.6      & 55.6669  & 3.6       & -0.073841 & 2      & 20    \\
O    & -76.5312  & 3.4      & 76.5313  & 3.4       & -0.074230 & 2      & 26    \\
F    & -101.3529 & 2.9      & 101.3528 & 2.9       & -0.074559 & 2      & 50    \\
Ne   & -130.3168 & 2.2      & 130.3174 & 2.2       & -0.074810 & 1      & 13    \\
Na   & -163.5942 & 1.8      & 163.5935 & 1.8       & -0.075061 & 1      & 13    \\
Mg   & -201.3471 & 1.6      & 201.3458 & 1.6       & -0.075283 & 1      & 13    \\
Al   & -243.7274 & 1.4      & 243.7251 & 1.4       & -0.075483 & 1      & 13    \\
Si   & -290.8789 & 1.3      & 290.8752 & 1.3       & -0.075665 & 1      & 13    \\
P    & -342.9386 & 1.2      & 342.9332 & 1.2       & -0.075832 & 1      & 12    \\
S    & -400.0373 & 1.1      & 400.0296 & 1.1       & -0.075986 & 1      & 12    \\
Cl   & -462.3002 & 1.1      & 462.2895 & 1.1       & -0.076129 & 1      & 12    \\
Ar   & -529.8474 & 1.0      & 529.8330 & 1.0       & -0.076262 & 2      & 19    \\
\hline
Mean     &           & 3.9      &          & 3.9       &           &        & 17   \\
     \hline
\hline
\end{tabular}
\end{table}

\begin{table}[h]
\caption{Energies, kinetic energies, chemical potentials and percentage errors relative to Kohn-Sham LDA exchange-only calculations using the $\gamma$TF$\lambda$vW(1.0, 1/5) functional. The number of uniques solutions generated from 25 random starting guesses is given, along with the number of iterations required for convergence.}
\begin{tabular}{lrrrrrrr}
\hline
\hline
Atom & E / $E_\text{h}$   & E \% Err & KE / $E_\text{h}$       & KE \% Err & $\mu$ / $E_\text{h}$        & $N_\text{sol}$ & $N_\text{iter}$ \\
\hline
H    & -0.5666   & 24.0     & 0.5666   & 23.7      & -0.065592 & 2      & 9     \\
He   & -2.8184   & 3.5      & 2.8184   & 3.4       & -0.070318 & 1      & 11    \\
Li   & -7.3227   & 1.8      & 7.3228   & 1.8       & -0.072356 & 1      & 12    \\
Be   & -14.4841  & 1.8      & 14.4842  & 1.8       & -0.073589 & 1      & 12    \\
B    & -24.6284  & 2.3      & 24.6285  & 2.3       & -0.074467 & 3      & 11    \\
C    & -38.0332  & 2.5      & 38.0334  & 2.5       & -0.075116 & 2      & 17    \\
N    & -54.9428  & 2.3      & 54.9430  & 2.3       & -0.075631 & 2      & 14    \\
O    & -75.5765  & 2.1      & 75.5766  & 2.1       & -0.076055 & 2      & 20    \\
F    & -100.1344 & 1.7      & 100.1344 & 1.7       & -0.076387 & 1      & 25    \\
Ne   & -128.8012 & 1.0      & 128.8010 & 1.0       & -0.076700 & 1      & 32    \\
Na   & -161.7488 & 0.7      & 161.7483 & 0.7       & -0.076965 & 1      & 15    \\
Mg   & -199.1382 & 0.4      & 199.1372 & 0.4       & -0.077205 & 1      & 17    \\
Al   & -241.1214 & 0.3      & 241.1196 & 0.3       & -0.077421 & 1      & 12    \\
Si   & -287.8423 & 0.2      & 287.8393 & 0.2       & -0.077618 & 1      & 12    \\
P    & -339.4376 & 0.2      & 339.4332 & 0.2       & -0.077797 & 2      & 11    \\
S    & -396.0381 & 0.1      & 396.0318 & 0.1       & -0.077959 & 1      & 13    \\
Cl   & -457.7689 & 0.1      & 457.7602 & 0.1       & -0.078116 & 2      & 13    \\
Ar   & -524.7502 & 0.0      & 524.7383 & 0.0       & -0.078255 & 1      & 13    \\
\hline
Mean     &           & 2.5      &          & 2.5       &           &        & 15   \\
\hline
\hline
\end{tabular}
\end{table}

\begin{table}[h]
\caption{Energies, kinetic energies, chemical potentials and percentage errors relative to Kohn-Sham LDA exchange-only calculations using the $\gamma$TF$\lambda$vW(1.0, 5/9) functional. The number of uniques solutions generated from 25 random starting guesses is given, along with the number of iterations required for convergence.}
\begin{tabular}{lrrrrrrr}
\hline
\hline
Atom & E / $E_\text{h}$   & E \% Err & KE / $E_\text{h}$       & KE \% Err & $\mu$ / $E_\text{h}$        & $N_\text{sol}$ & $N_\text{iter}$ \\
\hline
H    & -0.3700   & -19.0    & 0.3700   & -19.2     & -0.075668 & 1      & 9     \\
He   & -1.9801   & -27.3    & 1.9801   & -27.3     & -0.096059 & 1      & 10    \\
Li   & -5.3439   & -25.7    & 5.3439   & -25.7     & -0.105426 & 1      & 9     \\
Be   & -10.8387  & -23.8    & 10.8388  & -23.8     & -0.110809 & 2      & 10    \\
B    & -18.7736  & -22.0    & 18.7736  & -22.0     & -0.114368 & 3      & 11    \\
C    & -29.4141  & -20.7    & 29.4141  & -20.7     & -0.116948 & 1      & 10    \\
N    & -42.9953  & -19.9    & 42.9954  & -20.0     & -0.118938 & 1      & 13    \\
O    & -59.7294  & -19.3    & 59.7295  & -19.3     & -0.120541 & 1      & 21    \\
F    & -79.8105  & -19.0    & 79.8106  & -19.0     & -0.121877 & 1      & 21    \\
Ne   & -103.4184 & -18.9    & 103.4185 & -18.9     & -0.123015 & 1      & 31    \\
Na   & -130.7205 & -18.6    & 130.7206 & -18.6     & -0.124002 & 3      & 25    \\
Mg   & -161.8743 & -18.3    & 161.8743 & -18.3     & -0.124872 & 2      & 40    \\
Al   & -197.0283 & -18.0    & 197.0283 & -18.0     & -0.125646 & 1      & 18    \\
Si   & -236.3235 & -17.7    & 236.3234 & -17.7     & -0.126342 & 3      & 41    \\
P    & -279.8941 & -17.4    & 279.8939 & -17.4     & -0.126973 & 2      & 14    \\
S    & -327.8684 & -17.1    & 327.8681 & -17.1     & -0.127550 & 3      & 34    \\
Cl   & -380.3694 & -16.8    & 380.3689 & -16.8     & -0.128079 & 2      & 17    \\
Ar   & -437.5152 & -16.6    & 437.5145 & -16.6     & -0.128567 & 2      & 35    \\
\hline
Mean    &           & -19.8    &          & -19.8     &           &        & 21   \\
     \hline
\hline
\end{tabular}
\end{table}

\begin{table}[h]
\caption{Energies, kinetic energies, chemical potentials and percentage errors relative to Kohn-Sham LDA exchange-only calculations using the $\gamma$TF$\lambda$vW(1.0, 1.0) functional. The number of uniques solutions generated from 25 random starting guesses is given, along with the number of iterations required for convergence.}
\begin{tabular}{lrrrrrrr}
\hline
\hline
Atom & E / $E_\text{h}$   & E \% Err & KE / $E_\text{h}$       & KE \% Err & $\mu$ / $E_\text{h}$        & $N_\text{sol}$ & $N_\text{iter}$ \\
\hline
H    & -0.2618   & -42.7    & 0.2618   & -42.8     & -0.071453 & 1      & 14    \\
He   & -1.4774   & -45.8    & 1.4774   & -45.8     & -0.108237 & 2      & 10    \\
Li   & -4.1054   & -42.9    & 4.1054   & -42.9     & -0.130583 & 1      & 9     \\
Be   & -8.4922   & -40.3    & 8.4922   & -40.3     & -0.145309 & 1      & 10    \\
B    & -14.9258  & -38.0    & 14.9259  & -38.0     & -0.155638 & 1      & 10    \\
C    & -23.6568  & -36.3    & 23.6569  & -36.3     & -0.163252 & 1      & 10    \\
N    & -34.9084  & -35.0    & 34.9084  & -35.0     & -0.169098 & 1      & 10    \\
O    & -48.8831  & -33.9    & 48.8832  & -33.9     & -0.173740 & 1      & 11    \\
F    & -65.7674  & -33.2    & 65.7675  & -33.2     & -0.177530 & 1      & 11    \\
Ne   & -85.7343  & -32.8    & 85.7343  & -32.8     & -0.180697 & 1      & 11    \\
Na   & -108.9455 & -32.2    & 108.9455 & -32.2     & -0.183394 & 1      & 10    \\
Mg   & -135.5534 & -31.6    & 135.5534 & -31.6     & -0.185731 & 1      & 11    \\
Al   & -165.7021 & -31.1    & 165.7021 & -31.1     & -0.187782 & 1      & 11    \\
Si   & -199.5287 & -30.5    & 199.5287 & -30.5     & -0.189604 & 1      & 11    \\
P    & -237.1638 & -30.0    & 237.1638 & -30.0     & -0.191239 & 1      & 12    \\
S    & -278.7324 & -29.5    & 278.7324 & -29.5     & -0.192718 & 1      & 11    \\
Cl   & -324.3546 & -29.1    & 324.3546 & -29.1     & -0.194066 & 1      & 12    \\
Ar   & -374.1458 & -28.7    & 374.1457 & -28.7     & -0.195304 & 1      & 11    \\
\hline
Mean     &           & -34.6    &          & -34.6     &           &        & 11   \\
     \hline
\hline
\end{tabular}
\end{table}

\begin{table}[h]
\caption{Energies, kinetic energies, chemical potentials and percentage errors relative to Kohn-Sham LDA exchange-only calculations using the $\gamma$TF$\lambda$vW(0.697, 0.599) functional. The number of uniques solutions generated from 25 random starting guesses is given, along with the number of iterations required for convergence.}
\begin{tabular}{lrrrrrrr}
\hline
\hline
Atom & E / $E_\text{h}$   & E \% Err & KE / $E_\text{h}$       & KE \% Err & $\mu$ / $E_\text{h}$        & $N_\text{sol}$ & $N_\text{iter}$ \\
\hline
H    & -0.4137   & -9.5     & 0.4137   & -9.7      & -0.105483 & 2      & 8     \\
He   & -2.3020   & -15.5    & 2.3020   & -15.5     & -0.152349 & 2      & 10    \\
Li   & -6.3463   & -11.8    & 6.3463   & -11.8     & -0.178720 & 1      & 9     \\
Be   & -13.0570  & -8.2     & 13.0569  & -8.2      & -0.195291 & 1      & 9     \\
B    & -22.8565  & -5.0     & 22.8565  & -5.0      & -0.206586 & 1      & 10    \\
C    & -36.1110  & -2.7     & 36.1109  & -2.7      & -0.214783 & 1      & 10    \\
N    & -53.1460  & -1.0     & 53.1459  & -1.1      & -0.221033 & 1      & 10    \\
O    & -74.2566  & 0.3      & 74.2566  & 0.3       & -0.225987 & 1      & 10    \\
F    & -99.7138  & 1.3      & 99.7138  & 1.3       & -0.230039 & 1      & 11    \\
Ne   & -129.7688 & 1.8      & 129.7687 & 1.8       & -0.233438 & 1      & 11    \\
Na   & -164.6563 & 2.5      & 164.6562 & 2.5       & -0.236346 & 1      & 11    \\
Mg   & -204.5972 & 3.2      & 204.5971 & 3.2       & -0.238877 & 1      & 10    \\
Al   & -249.8005 & 3.9      & 249.8003 & 3.9       & -0.241111 & 1      & 11    \\
Si   & -300.4644 & 4.6      & 300.4641 & 4.6       & -0.243104 & 1      & 11    \\
P    & -356.7781 & 5.3      & 356.7777 & 5.3       & -0.244899 & 1      & 11    \\
S    & -418.9225 & 5.9      & 418.9219 & 5.9       & -0.246530 & 1      & 11    \\
Cl   & -487.0711 & 6.5      & 487.0702 & 6.5       & -0.248022 & 1      & 11    \\
Ar   & -561.3907 & 7.0      & 561.3896 & 7.0       & -0.249395 & 1      & 11    \\
\hline
Mean     &           & -0.6     &          & -0.6      &           &        & 10   \\
     \hline
\hline
\end{tabular}
\end{table}

\begin{table}[h]
\caption{Energies, kinetic energies, chemical potentials and percentage errors relative to Kohn-Sham LDA exchange-only calculations using the OL1 functional. The number of uniques solutions generated from 25 random starting guesses is given, along with the number of iterations required for convergence.}
\begin{tabular}{lrrrrrrr}
\hline
\hline
Atom & E / $E_\text{h}$   & E \% Err & KE / $E_\text{h}$       & KE \% Err & $\mu$ / $E_\text{h}$        & $N_\text{sol}$ & $N_\text{iter}$ \\
\hline
H    & -0.6571   & 43.8     & 0.6570   & 43.5      & -0.058531 & 1      & 9     \\
He   & -3.1806   & 16.8     & 3.1805   & 16.7      & -0.060935 & 1      & 12    \\
Li   & -8.1483   & 13.3     & 8.1483   & 13.3      & -0.061956 & 2      & 10    \\
Be   & -15.9682  & 12.3     & 15.9684  & 12.3      & -0.062561 & 1      & 11    \\
B    & -26.9684  & 12.1     & 26.9687  & 12.0      & -0.062948 & 1      & 13    \\
C    & -41.4275  & 11.6     & 41.4277  & 11.6      & -0.063261 & 2      & 15    \\
N    & -59.5908  & 11.0     & 59.5914  & 10.9      & -0.063495 & 2      & 13    \\
O    & -81.6776  & 10.4     & 81.6775  & 10.4      & -0.063709 & 1      & 15    \\
F    & -107.8892 & 9.6      & 107.8885 & 9.6       & -0.063890 & 1      & 12    \\
Ne   & -138.4102 & 8.6      & 138.4085 & 8.6       & -0.064053 & 2      & 45    \\
Na   & -173.4126 & 7.9      & 173.4094 & 7.9       & -0.064201 & 2      & 13    \\
Mg   & -213.0578 & 7.5      & 213.0524 & 7.5       & -0.064331 & 1      & 21    \\
Al   & -257.4977 & 7.1      & 257.4891 & 7.1       & -0.064469 & 2      & 32    \\
Si   & -306.8762 & 6.9      & 306.8633 & 6.9       & -0.064590 & 2      & 31    \\
P    & -361.3303 & 6.6      & 361.3117 & 6.6       & -0.064691 & 1      & 18    \\
S    & -420.9904 & 6.4      & 420.9645 & 6.4       & -0.064816 & 2      & 13    \\
Cl   & -485.9819 & 6.3      & 485.9464 & 6.3       & -0.064900 & 1      & 13    \\
Ar   & -556.4245 & 6.1      & 556.3772 & 6.1       & -0.064998 & 1      & 12    \\
\hline
Mean     &           & 11.3     &          & 11.3      &           &        & 17   \\
     \hline
\hline
\end{tabular}
\end{table}

\begin{table}[h]
\caption{Energies, kinetic energies, chemical potentials and percentage errors relative to Kohn-Sham LDA exchange-only calculations using the P92 functional. The number of uniques solutions generated from 25 random starting guesses is given, along with the number of iterations required for convergence.}
\begin{tabular}{lrrrrrrr}
\hline
\hline
Atom & E / $E_\text{h}$   & E \% Err & KE / $E_\text{h}$       & KE \% Err & $\mu$ / $E_\text{h}$        & $N_\text{sol}$ & $N_\text{iter}$ \\
\hline
H    & -0.6661   & 45.8     & 0.6660   & 45.5      & -0.059172 & 1      & 9     \\
He   & -3.2216   & 18.3     & 3.2214   & 18.2      & -0.061545 & 1      & 10    \\
Li   & -8.2486   & 14.7     & 8.2486   & 14.7      & -0.062545 & 1      & 11    \\
Be   & -16.1580  & 13.6     & 16.1581  & 13.6      & -0.063135 & 1      & 10    \\
B    & -27.2795  & 13.3     & 27.2796  & 13.3      & -0.063520 & 1      & 12    \\
C    & -41.8934  & 12.9     & 41.8936  & 12.9      & -0.063807 & 1      & 29    \\
N    & -60.2460  & 12.2     & 60.2461  & 12.2      & -0.064030 & 1      & 12    \\
O    & -82.5583  & 11.6     & 82.5580  & 11.6      & -0.064226 & 1      & 10    \\
F    & -109.0324 & 10.7     & 109.0334 & 10.7      & -0.064395 & 1      & 17    \\
Ne   & -139.8522 & 9.7      & 139.8502 & 9.7       & -0.064556 & 1      & 12    \\
Na   & -175.1923 & 9.1      & 175.1886 & 9.1       & -0.064691 & 1      & 15    \\
Mg   & -215.2140 & 8.6      & 215.2078 & 8.6       & -0.064826 & 1      & 39    \\
Al   & -260.0699 & 8.2      & 260.0602 & 8.2       & -0.064949 & 2      & 12    \\
Si   & -309.9045 & 7.9      & 309.8901 & 7.9       & -0.065064 & 2      & 18    \\
P    & -364.8551 & 7.7      & 364.8345 & 7.7       & -0.065162 & 1      & 13    \\
S    & -425.0530 & 7.5      & 425.0241 & 7.5       & -0.065278 & 2      & 37    \\
Cl   & -490.6235 & 7.3      & 490.5842 & 7.3       & -0.065361 & 1      & 13    \\
Ar   & -561.6873 & 7.1      & 561.6348 & 7.1       & -0.065455 & 1      & 15    \\
\hline
Mean     &           & 12.6     &          & 12.5      &           &        & 16   \\
     \hline
\hline
\end{tabular}
\end{table}

\begin{table}[h]
\caption{Energies, kinetic energies, chemical potentials and percentage errors relative to Kohn-Sham LDA exchange-only calculations using the E00 functional. The number of uniques solutions generated from 25 random starting guesses is given, along with the number of iterations required for convergence.}
\begin{tabular}{lrrrrrrr}
\hline
\hline
Atom & E / $E_\text{h}$   & E \% Err & KE / $E_\text{h}$       & KE \% Err & $\mu$ / $E_\text{h}$        & $N_\text{sol}$ & $N_\text{iter}$ \\
\hline
H    & -0.6214   & 36.0     & 0.6214   & 35.7      & -0.089637 & 1      & 11    \\
He   & -3.0901   & 13.5     & 3.0901   & 13.4      & -0.092600 & 2      & 10    \\
Li   & -7.9888   & 11.1     & 7.9888   & 11.1      & -0.093518 & 2      & 13    \\
Be   & -15.7299  & 10.6     & 15.7299  & 10.6      & -0.094093 & 1      & 11    \\
B    & -26.6444  & 10.7     & 26.6444  & 10.7      & -0.094514 & 1      & 10    \\
C    & -41.0135  & 10.5     & 41.0133  & 10.5      & -0.094844 & 2      & 12    \\
N    & -59.0839  & 10.0     & 59.0835  & 10.0      & -0.095113 & 1      & 11    \\
O    & -81.0770  & 9.6      & 81.0764  & 9.6       & -0.095336 & 1      & 12    \\
F    & -107.1948 & 8.9      & 107.1937 & 8.8       & -0.095527 & 1      & 12    \\
Ne   & -137.6233 & 7.9      & 137.6214 & 7.9       & -0.095691 & 1      & 11    \\
Na   & -172.5355 & 7.4      & 172.5325 & 7.4       & -0.095836 & 1      & 11    \\
Mg   & -212.0935 & 7.0      & 212.0890 & 7.0       & -0.095963 & 1      & 12    \\
Al   & -256.4502 & 6.7      & 256.4435 & 6.7       & -0.096077 & 2      & 14    \\
Si   & -305.7502 & 6.5      & 305.7406 & 6.5       & -0.096179 & 1      & 11    \\
P    & -360.1310 & 6.3      & 360.1176 & 6.3       & -0.096272 & 1      & 12    \\
S    & -419.7241 & 6.1      & 419.7057 & 6.1       & -0.096356 & 1      & 12    \\
Cl   & -484.6550 & 6.0      & 484.6302 & 6.0       & -0.096433 & 1      & 12    \\
Ar   & -555.0443 & 5.8      & 555.0116 & 5.8       & -0.096503 & 1      & 12    \\
\hline
Mean     &           & 10.0     &          & 10.0      &           &        & 12   \\
     \hline
\hline
\end{tabular}
\end{table}

\begin{table}[h]
\caption{Energies, kinetic energies, chemical potentials and percentage errors relative to Kohn-Sham LDA exchange-only calculations using the conjB96a functional. The number of uniques solutions generated from 25 random starting guesses is given, along with the number of iterations required for convergence.}
\begin{tabular}{lrrrrrrr}
\hline
\hline
Atom & E / $E_\text{h}$   & E \% Err & KE / $E_\text{h}$       & KE \% Err & $\mu$ / $E_\text{h}$        & $N_\text{sol}$ & $N_\text{iter}$ \\
\hline
H    & -0.6985   & 52.8     & 0.5237   & 14.4      & -0.048231 & 22     & 107   \\
He   & -3.3611   & 23.4     & 2.7042   & -0.8      & -0.050571 & 24     & 165   \\
Li   & -8.6605   & 20.4     & 6.7706   & -5.9      & -0.049912 & 24     & 12    \\
Be   & -16.7814  & 18.0     & 13.5983  & -4.4      & -0.048358 & 18     & 12    \\
B    & -28.1660  & 17.0     & 22.9255  & -4.8      & -0.048059 & 17     & 11    \\
C    & -43.0552  & 16.0     & 36.4574  & -1.8      & -0.046907 & 14     & 13    \\
N    & -61.6855  & 14.9     & 52.1672  & -2.9      & -0.046364 & 15     & 11    \\
O    & -84.2716  & 13.9     & 74.1021  & 0.1       & -0.045681 & 14     & 13    \\
F    & -111.0130 & 12.7     & 98.0841  & -0.4      & -0.044106 & 12     & 14    \\
Ne   & -142.4269 & 11.7     & 125.3870 & -1.7      & -0.043774 & 13     & 20    \\
Na   & -178.0975 & 10.9     & 143.7221 & -10.5     & -0.044464 & 11     & 17    \\
Mg   & -218.4343 & 10.2     & 197.9070 & -0.2      & -0.042907 & 10     & 12    \\
Al   & -262.5384 & 9.2      & 225.1514 & -6.3      & -0.042794 & 13     & 14    \\
Si   & -313.6979 & 9.2      & 282.3623 & -1.7      & -0.042488 & 11     & 22    \\
P    & -368.9056 & 8.9      & 324.6542 & -4.2      & -0.042409 & 11     & 17    \\
S    & -429.3434 & 8.6      & 384.0853 & -2.9      & -0.042214 & 9      & 30    \\
Cl   & -495.1424 & 8.3      & 441.6746 & -3.4      & -0.042190 & 10     & 24    \\
Ar   & -566.4335 & 8.0      & 503.8672 & -3.9      & -0.042219 & 10     & 24    \\
\hline
 Mean    &           & 15.2     &          & -2.3      &           &        & 30   \\
     \hline
\hline
\end{tabular}
\end{table}

\begin{table}[h]
\caption{Energies, kinetic energies, chemical potentials and percentage errors relative to Kohn-Sham LDA exchange-only calculations using the conjPW91 functional. The number of uniques solutions generated from 25 random starting guesses is given, along with the number of iterations required for convergence.}
\begin{tabular}{lrrrrrrr}
\hline
\hline
Atom & E / $E_\text{h}$   & E \% Err & KE / $E_\text{h}$       & KE \% Err & $\mu$ / $E_\text{h}$        & $N_\text{sol}$ & $N_\text{iter}$ \\
\hline
H    & -0.9203   & 101.4    & 0.7482   & 63.4      & -0.049104 & 25     & 67    \\
He   & -4.0764   & 49.7     & 3.3070   & 21.3      & -0.049938 & 24     & 48    \\
Li   & -9.7505   & 35.5     & 7.9646   & 10.7      & -0.052354 & 24     & 18    \\
Be   & -18.5437  & 30.4     & 15.7978  & 11.1      & -0.048754 & 20     & 20    \\
B    & -30.5548  & 27.0     & 26.4095  & 9.7       & -0.047947 & 22     & 22    \\
C    & -46.6498  & 25.7     & 36.1197  & -2.7      & -0.047053 & 21     & 19    \\
N    & -66.1488  & 23.2     & 54.1167  & 0.8       & -0.045243 & 20     & 76    \\
O    & -89.5614  & 21.0     & 76.5623  & 3.5       & -0.045377 & 20     & 47    \\
F    & -117.1302 & 18.9     & 102.7915 & 4.4       & -0.044315 & 14     & 16    \\
Ne   & -149.1257 & 17.0     & 121.3869 & -4.8      & -0.043951 & 15     & 16    \\
Na   & -185.5806 & 15.5     & 156.2903 & -2.7      & -0.043360 & 12     & 14    \\
Mg   & -226.6356 & 14.3     & 190.0676 & -4.1      & -0.042817 & 10     & 21    \\
Al   & -272.4251 & 13.3     & 227.5450 & -5.3      & -0.042306 & 15     & 18    \\
Si   & -323.0794 & 12.5     & 277.4811 & -3.4      & -0.042189 & 13     & 18    \\
P    & -376.9933 & 11.2     & 314.6017 & -7.2      & -0.041719 & 15     & 17    \\
S    & -439.4825 & 11.1     & 380.7593 & -3.7      & -0.044956 & 11     & 18    \\
Cl   & -505.4718 & 10.5     & 443.3154 & -3.1      & -0.042053 & 10     & 32    \\
Ar   & -577.5859 & 10.1     & 533.5227 & 1.7       & -0.041757 & 7      & 38    \\
\hline
Mean     &           & 24.9     &          & 5.0       &           &        & 29   \\
     \hline
\hline
\end{tabular}
\end{table}

\clearpage
\newpage
\section{Molecular Data}

All calculations were carried out at the geometries given in [ A. M. Teale, O. B. Lutn{\ae}s, T. Helgaker, D. J. Tozer, and J. Gauss, {\emph J. Chem. Phys.} {\bf 138}, 024111 (2013) ], using an uncontracted basis set consisting of the primitive exponents from the def2-qzvp basis set. All calculations use a superposition-of-atomic densities initial guess constructed from atomic densities calculated with the same functional in the same basis set.

\begin{table}[h]
\caption{Energies, kinetic energies, chemical potentials and percentage errors relative to Kohn-Sham LDA exchange-only calculations using the $\gamma$TF$\lambda$vW(1.0, 1/5) functional. The error in the particle number at convergence and the number of TRIM iterations required to reach convergence are also given.}
\begin{tabular}{lrrrrrrrr}
\hline
\hline
Molecule   & E / $E_\text{h}$          & \%Err E & T / $E_\text{h}$       & \%Err T & $\mu$ / $E_\text{h}$         & \%Err $\mu$ & $N_\text{err}$        & $N_\text{iter}$  \\
\hline
AlF   & -340.6465  & 0.5     & 343.0040 & 1.2     & -0.0562    & -70.9    & 5.38$\times 10^{-10}$    & 31   \\
C$_2$H$_4$  & -78.0349   & 1.5     & 80.0810  & 3.9     & -0.0802    & -62.1    & 1.00$\times 10^{-12}$    & 14   \\
C$_2$H$_4$O & -152.3416 & 0.9     & 158.7265 & 4.9     & -0.0561 & -70.0    & 9.80$\times 10^{-11}$     & 83   \\
C$_3$H$_4$  & -115.9664  & 1.6     & 118.8266 & 3.9     & -0.0812    & -55.7    & 1.00$\times 10^{-12}$    & 14   \\
CH$_2$O  & -113.4265  & 0.7     & 118.0374 & 4.7     & -0.0438    & -76.3    & 6.00$\times 10^{-12}$       & 17   \\
CH$_3$F  & -138.8203  & 0.9     & 142.9639 & 3.8     & -0.0442    & -82.3    & -2.21$\times 10^{-08}$ & 875  \\
CH$_4$   & -40.0965   & 1.4     & 41.4599  & 4.5     & -0.0787    & -73.9    & 2.00$\times 10^{-12}$       & 14   \\
CO    & -112.1664  & 0.6     & 116.6534 & 4.6     & -0.0415    & -85.6    & 1.00$\times 10^{-12}$       & 16   \\
FCCH  & -175.4791  & 1.0     & 180.3629 & 3.7     & -0.0595    & -72.6    & 2.50$\times 10^{-11}$     & 20   \\
FCN   & -192.6828  & 1.5     & 195.0991 & 2.8     & -0.0765    & -72.7    & 1.00$\times 10^{-12}$    & 14   \\
H$_2$C$_2$O & -151.2381  & 0.8     & 156.8709 & 4.5     & -0.0596    & -66.5    & 1.00$\times 10^{-12}$    & 18   \\
H$_2$O   & -76.4396 & 1.6     & 77.7179  & 3.1     & -0.0776 & -65.0    & 1.00$\times 10^{-12}$    & 65   \\
H$_2$S   & -396.4350  & -0.1    & 398.6772 & 0.5     & -0.0339    & -82.3    & 5.30$\times 10^{-11}$     & 25   \\
HCN   & -93.2387   & 1.6     & 94.9055  & 3.4     & -0.0790    & -72.8    & 1.00$\times 10^{-12}$    & 17   \\
HCP   & -377.9173 & 0.3     & 378.9038 & 0.5     & -0.0799 & -63.8    & 1.00$\times 10^{-12}$    & 17   \\
HFCO  & -213.8787  & 1.5     & 216.5027 & 2.7     & -0.0784    & -68.0    & 1.00$\times 10^{-12}$    & 15   \\
HF    & -100.5158  & 1.4     & 101.3235 & 2.1     & -0.0742    & -75.9    & 1.00$\times 10^{-12}$    & 19   \\
HOF   & -176.0344  & 1.6     & 177.5135 & 2.5     & -0.0727    & -67.2    & -1.00$\times 10^{-12}$      & 26   \\
LiF   & -107.1059  & 1.2     & 108.3103 & 2.3     & -0.0553    & -70.5    & -1.22$\times 10^{-8}$ & 31   \\
LiH   & -7.9029    & 2.6     & 7.9200   & 2.6     & -0.0749    & -41.6    & -3.00$\times 10^{-12}$      & 12   \\
N$_2$    & -109.5191  & 1.6     & 111.3462 & 3.3     & -0.0568    & -82.8    & 1.00$\times 10^{-12}$    & 13   \\
N$_2$O   & -183.7554  & 1.0     & 189.8941 & 4.4     & -0.0427    & -84.0    & 1.00$\times 10^{-12}$    & 22   \\
NH$_3$   & -56.3694   & 1.6     & 57.8347  & 4.0     & -0.0783    & -57.7    & 1.00$\times 10^{-12}$    & 21   \\
O$_3$    & -225.8516  & 1.7     & 230.2807 & 3.1     & -0.0241    & -80.9    & 1.32$\times 10^{-10}$    & 15   \\
OCS   & -507.5494  & 0.1     & 514.5887 & 1.4     & -0.0534    & -77.2    & 2.00$\times 10^{-12}$       & 42   \\
OF$_2$   & -275.5355  & 1.6     & 277.8010 & 2.4     & -0.0005    & -99.8    & 1.00$\times 10^{-12}$    & 15   \\
PN    & -394.2803  & 0.4     & 395.1123 & 0.6     & -0.0788    & -67.6    & 1.00$\times 10^{-12}$    & 23   \\
SO$_2$   & -546.8654  & 0.5     & 549.1974 & 1.0     & -0.0749    & -70.6    & 1.00$\times 10^{-12}$    & 35   \\
\hline
 Mean     &       & 1.1     &     & 2.9     &     & -72.0    &         & 55   \\
 Max     &            &         &          &         &            &          &          & 875  \\
 Min     &            &         &          &         &            &          &          & 12   \\
 Median     &            &         &          &         &            &          &       & 19 \\
      \hline
\hline
\end{tabular}
\end{table}

\begin{table}[h]
\caption{Energies, kinetic energies, chemical potentials and percentage errors relative to Kohn-Sham LDA exchange-only calculations using the $\gamma$TF$\lambda$vW(0.697, 0.599) functional. The error in the particle number at convergence and the number of TRIM iterations required to reach convergence are also given.}
\begin{tabular}{lrrrrrrrr}
\hline
\hline
Molecule   & E / $E_\text{h}$          & \%Err E & T / $E_\text{h}$       & \%Err T & $\mu$ / $E_\text{h}$         & \%Err $\mu$ & $N_\text{err}$        & $N_\text{iter}$  \\
\hline
AlF   & -348.9597  & 2.9     & 351.2304 & 3.6     & -0.1671    & -13.5    & 2.20$\times 10^{-8}$ & 163 \\
C$_2$H$_4$  & -74.8898   & -2.6    & 74.0316  & -3.9    & -0.1896    & -10.4    & 1.00$\times 10^{-12}$   & 14  \\
C$_2$H$_4$O & -149.6016  & -1.0    & 147.7657 & -2.3    & -0.2047    & 9.5      & 1.00$\times 10^{-12}$   & 13  \\
C$_3$H$_4$  & -111.3867  & -2.5    & 109.7341 & -4.1    & -0.1982    & 8.1      & 1.00$\times 10^{-12}$   & 14  \\
CH$_2$O  & -110.0981  & -2.2    & 114.7398 & 1.8     & -0.1220    & -34.0    & 1.00$\times 10^{-12}$   & 18  \\
CH$_3$F  & -137.9407  & 0.3     & 137.0449 & -0.5    & -0.2051    & -17.7    & 1.00$\times 10^{-12}$   & 18  \\
CH$_4$   & -38.4600   & -2.7    & 38.2221  & -3.7    & -0.1749    & -42.1    & -2.00$\times 10^{-12}$     & 13  \\
CO    & -108.6899  & -2.5    & 114.1130 & 2.3     & -0.1116    & -61.2    & 1.00$\times 10^{-12}$   & 16  \\
FCCH  & -171.7326 & -1.2    & 175.1257 & 0.7     & -0.1618 & -25.6    & 1.00$\times 10^{-12}$   & 13  \\
FCN   & -188.1848  & -0.9    & 191.7341 & 1.0     & -0.1690    & -39.7    & 1.00$\times 10^{-12}$   & 15  \\
H$_2$C$_2$O & -146.4618  & -2.3    & 150.7203 & 0.4     & -0.1527    & -14.1    & 1.00$\times 10^{-12}$   & 14  \\
H$_2$O   & -75.4345   & 0.2     & 75.6204  & 0.4     & -0.1991    & -10.1    & 1.00$\times 10^{-12}$   & 14  \\
H$_2$S   & -420.1038  & 5.9     & 419.7592 & 5.8     & -0.1979    & 3.7      & 1.00$\times 10^{-12}$   & 26  \\
HCN   & -90.2368   & -1.6    & 89.6496  & -2.4    & -0.2241    & -22.8    & 1.00$\times 10^{-12}$   & 15  \\
HCP   & -393.7490  & 4.5     & 393.0424 & 4.3     & -0.2338    & 5.8      & -1.00$\times 10^{-12}$     & 16  \\
HFCO  & -208.1342  & -1.2    & 216.5591 & 2.7     & -0.1244    & -49.2    & 1.00$\times 10^{-12}$   & 24  \\
HF    & -100.2943  & 1.2     & 100.4806 & 1.3     & -0.2113    & -31.2    & 1.00$\times 10^{-12}$   & 17  \\
HOF   & -174.8905  & 1.0     & 174.2761 & 0.6     & -0.2340    & 5.5      & -1.00$\times 10^{-12}$     & 15  \\
LiF   & -105.6279  & -0.2    & 107.1768 & 1.2     & -0.1387    & -26.1    & 1.00$\times 10^{-12}$   & 629 \\
LiH   & -6.8511    & -11.1   & 6.7130   & -13.1   & -0.1643    & 28.0     & 4.00$\times 10^{-12}$      & 13  \\
N$_2$    & -106.7333  & -0.9    & 106.2271 & -1.5    & -0.2462    & -25.5    & 1.00$\times 10^{-12}$   & 12  \\
N$_2$O   & -179.4690  & -1.3    & 184.0201 & 1.1     & -0.1591    & -40.2    & 1.00$\times 10^{-12}$   & 15  \\
NH$_3$   & -54.9201   & -1.0    & 54.9628  & -1.2    & -0.1868    & 1.0      & 1.00$\times 10^{-12}$   & 25  \\
O$_3$    & -223.7071  & 0.7     & 222.5911 & -0.4    & -0.2534    & 100.1    & 1.00$\times 10^{-12}$   & 12  \\
OCS   & -526.9794  & 3.9     & 535.0718 & 5.5     & -0.1067    & -54.3    & 1.50$\times 10^{-11}$    & 71  \\
OF$_2$   & -274.3525  & 1.2     & 273.0053 & 0.7     & -0.2413    & 3.9      & 1.00$\times 10^{-12}$   & 13  \\
PN    & -410.2349  & 4.4     & 409.5116 & 4.3     & -0.2589    & 6.5      & 1.00$\times 10^{-12}$   & 17  \\
SO$_2$   & -568.1559 & 4.4     & 566.5454 & 4.1     & -0.2679 & 5.2      & 1.00$\times 10^{-12}$   & 19  \\
\hline
Mean      &     & -0.2    &    & 0.3     &     & -12.2    &       & 45  \\
Max      &            &         &          &         &            &          &        & 629 \\
Min      &            &         &          &         &            &          &      & 12  \\
Median      &            &         &          &         &            &          &    & 15 \\
      \hline
\hline
\end{tabular}
\end{table}

\begin{table}[h]
\caption{Energies, kinetic energies, chemical potentials and percentage errors relative to Kohn-Sham LDA exchange-only calculations using the OL1 functional. The error in the particle number at convergence and the number of TRIM iterations required to reach convergence are also given.}
\begin{tabular}{lrrrrrrrr}
\hline
\hline
Molecule   & E / $E_\text{h}$          & \%Err E & T / $E_\text{h}$       & \%Err T & $\mu$ / $E_\text{h}$         & \%Err $\mu$ & $N_\text{err}$        & $N_\text{iter}$  \\
\hline
AlF   & -364.9043  & 7.6     & 366.8633   & 8.2     & -0.0508    & -73.7    & -2.00$\times 10^{-12}$      & 32  \\
C$_2$H$_4$  & -85.0226   & 10.6    & 87.2866    & 13.3    & -0.0670    & -68.3    & -2.40$\times 10^{-11}$    & 15  \\
C$_2$H$_4$O & -165.7064 & 9.7     & 171.522536 & 13.4    & -0.0480 & -74.3    & 1.00$\times 10^{-12}$    & 157 \\
C$_3$H$_4$  & -126.2810  & 10.6    & 129.5012   & 13.2    & -0.0675    & -63.2    & 4.50$\times 10^{-11}$     & 15  \\
CH$_2$O  & -123.2826  & 9.5     & 127.2965   & 12.9    & -0.0431    & -76.7    & 1.64$\times 10^{-8}$  & 133 \\
CH$_3$F  & -150.3454  & 9.3     & 154.0054   & 11.8    & -0.0424    & -83.0    & 1.05$\times 10^{-7}$ & 108 \\
CH$_4$   & -43.7453   & 10.7    & 45.2149    & 14.0    & -0.0660    & -78.1    & -1.00$\times 10^{-12}$      & 21  \\
CO    & -121.9212  & 9.3     & 125.7164   & 12.7    & -0.0424    & -85.2    & 3.30$\times 10^{-11}$     & 20  \\
FCCH  & -190.8285  & 9.8     & 193.5515   & 11.3    & -0.0668    & -69.3    & 8.00$\times 10^{-12}$       & 20  \\
FCN   & -207.6389  & 9.4     & 212.3512   & 11.8    & -0.0389    & -86.1    & 2.00$\times 10^{-12}$       & 86  \\
H$_2$C$_2$O & -164.4404  & 9.6     & 169.5801   & 13.0    & -0.0539    & -69.7    & -1.00$\times 10^{-12}$      & 847 \\
H$_2$O   & -81.9056   & 8.8     & 84.8094    & 12.5    & -0.0135    & -93.9    & 5.80$\times 10^{-11}$     & 39  \\
H$_2$S   & -422.2108  & 6.4     & 422.9134   & 6.6     & -0.0659    & -65.4    & -2.00$\times 10^{-12}$      & 24  \\
HCN   & -101.2692  & 10.4    & 103.0938   & 12.3    & -0.0661    & -77.2    & 1.00$\times 10^{-12}$    & 17  \\
HCP   & -403.2243  & 7.0     & 404.3778   & 7.3     & -0.0668    & -69.8    & -5.00$\times 10^{-12}$      & 24  \\
HFCO  & -229.7714  & 9.0     & 236.2250   & 12.0    & -0.0453    & -81.5    & -2.00$\times 10^{-11}$      & 83  \\
HF    & -108.3247  & 9.3     & 109.1474   & 10.0    & -0.0640    & -79.2    & 1.00$\times 10^{-12}$    & 45  \\
HOF   & -189.5247  & 9.4     & 191.8027   & 10.7    & -0.0590    & -73.4    & 8.20$\times 10^{-11}$     & 24  \\
LiF   & -115.7611 & 9.3     & 116.731066 & 10.3    & -0.0514 & -72.6    & 4.05$\times 10^{-08}$  & 962 \\
LiH   & -8.8066    & 14.3    & 8.8394     & 14.5    & -0.0639    & -50.3    & -1.11$\times 10^{-10}$   & 16  \\
N$_2$    & -118.7276  & 10.2    & 120.6906   & 12.0    & -0.0482    & -85.4    & 1.00$\times 10^{-12}$    & 14  \\
N$_2$O   & -200.0901  & 10.0    & 203.5463   & 11.9    & -0.0575    & -78.4    & 1.00$\times 10^{-12}$    & 18  \\
NH$_3$   & -61.1980   & 10.3    & 62.7426    & 12.8    & -0.0660    & -64.3    & 1.00$\times 10^{-12}$    & 22  \\
O$_3$    & -243.9627  & 9.8     & 248.7096   & 11.3    & -0.0176    & -86.1    & 1.00$\times 10^{-12}$    & 15  \\
OCS   & -542.4220  & 6.9     & 548.3321   & 8.1     & -0.0486    & -79.2    & 9.00$\times 10^{-12}$       & 21  \\
OF$_2$   & -297.0216  & 9.5     & 299.4897   & 10.4    & 0.0044     & -101.9   & 1.55$\times 10^{-9}$    & 15  \\
PN    & -420.7682  & 7.1     & 421.7564   & 7.4     & -0.0662    & -72.8    & 1.48$\times 10^{-8}$  & 21  \\
SO$_2$   & -583.8832  & 7.3     & 586.5785   & 7.8     & -0.0630    & -75.3    & 1.00$\times 10^{-12}$    & 36  \\
\hline
Mean      &      & 9.3     &      & 11.2    &     & -76.2    &        & 102 \\
Max      &            &         &            &         &            &          &          & 962 \\
Min      &            &         &            &         &            &          &        & 14  \\
Median      &            &         &            &         &            &          &       & 23 \\
      \hline
\hline
\end{tabular}
\end{table}

\begin{table}[h]
\caption{Energies, kinetic energies, chemical potentials and percentage errors relative to Kohn-Sham LDA exchange-only calculations using the P92 functional. The error in the particle number at convergence and the number of TRIM iterations required to reach convergence are also given.}
\begin{tabular}{lrrrrrrrr}
\hline
\hline
Molecule   & E / $E_\text{h}$          & \%Err E & T / $E_\text{h}$       & \%Err T & $\mu$ / $E_\text{h}$         & \%Err $\mu$ & $N_\text{err}$        & $N_\text{iter}$  \\
\hline
AlF   & -368.6252     & 8.7     & 370.5631     & 9.3     & -0.0509     & -73.7    & 5.19$\times 10^{-10}$   & 20  \\
C$_2$H$_4$  & -85.9934     & 11.9    & 88.2365     & 14.5    & -0.0675     & -68.1    & 1.00$\times 10^{-12}$   & 15  \\
C$_2$H$_4$O & -167.5745     & 10.9    & 173.3042     & 14.6    & -0.0521     & -72.1    & -1.50$\times 10^{-11}$   & 16  \\
C$_3$H$_4$  & -127.7181     & 11.8    & 130.9113     & 14.4    & -0.0680      & -62.9    & 1.00$\times 10^{-12}$   & 15  \\
CH$_2$O  & -124.6590     & 10.7    & 128.6321     & 14.1    & -0.0438     & -76.3    & 2.00$\times 10^{-12}$      & 44  \\
CH$_3$F  & -151.9907     & 10.5    & 155.6110     & 13.0    & -0.0432     & -82.7    & 3.00$\times 10^{-12}$      & 280 \\
CH$_4$   & -44.2497      & 11.9    & 45.7043     & 15.2    & -0.0665     & -78.0    & 4.00$\times 10^{-12}$      & 22  \\
CO    & -123.2801 & 10.5    & 127.0385 & 13.9    & -0.0431 & -85.0    & 1.00$\times 10^{-12}$   & 209 \\
FCCH  & -192.3150     & 10.6    & 196.7229     & 13.1    & -0.0549     & -74.8    & 1.00$\times 10^{-12}$   & 20  \\
FCN   & -209.9789     & 10.6    & 214.4061     & 12.9    & -0.0504     & -82.0    & 1.00$\times 10^{-12}$   & 150 \\
H$_2$C$_2$O & -166.2842     & 10.9    & 171.3757     & 14.1    & -0.0544     & -69.4    & -2.10$\times 10^{-11}$   & 373 \\
H$_2$O   & -83.5565     & 11.0    & 84.8726     & 12.6    & -0.0665     & -70.0    & 1.00$\times 10^{-12}$   & 61  \\
H$_2$S   & -425.8352     & 7.3     & 427.6413     & 7.8     & -0.0365     & -80.9    & 1.74$\times 10^{-10}$   & 52  \\
HCN   & -102.4019      & 11.6    & 104.2116     & 13.5    & -0.0666     & -77.0    & 1.00$\times 10^{-12}$   & 16  \\
HCP   & -407.2264     & 8.1     & 408.3722     & 8.4     & -0.0673     & -69.6    & 1.00$\times 10^{-12}$   & 37  \\
HFCO  & -232.2891     & 10.2    & 238.6784     & 13.2    & -0.0460      & -81.2    & 1.00$\times 10^{-12}$   & 339 \\
HF    & -109.4777     & 10.4    & 110.2924     & 11.2    & -0.0645     & -79.0    & 1.00$\times 10^{-12}$   & 30  \\
HOF   & -191.9181     & 10.8    & 193.5264     & 11.7    & -0.0624     & -71.9    & 1.00$\times 10^{-12}$   & 30  \\
LiF   & -117.0071     & 10.5    & 117.9652     & 11.4    & -0.0519     & -72.3    & 1.41$\times 10^{-8}$ & 29  \\
LiH   & -8.9158     & 15.7    & 8.9476     & 15.9    & -0.0644     & -49.8    & -1.00$\times 10^{-11}$     & 17  \\
N$_2$    & -120.0409     & 11.4    & 121.9884     & 13.2    & -0.0492     & -85.1    & 1.00$\times 10^{-12}$   & 13  \\
N$_2$O   & -201.5260 & 10.8    & 207.0565 & 13.8    & -0.0424 & -84.1    & 1.00$\times 10^{-12}$   & 16  \\
NH$_3$   & -61.8830     & 11.5    & 63.4133     & 14.0    & -0.0664     & -64.1    & 1.00$\times 10^{-12}$      & 21  \\
O$_3$    & -246.6107     & 11.0    & 251.3217     & 12.5    & -0.0190      & -85.0    & 1.00$\times 10^{-12}$   & 15  \\
OCS   & -547.8524     & 8.0     & 553.7040     & 9.2     & -0.0491     & -79.0    & 3.51$\times 10^{-10}$   & 130 \\
OF$_2$   & -300.1893     & 10.7    & 302.6336     & 11.6    & 0.0029     & -101.2   & 3.52$\times 10^{-10}$   & 19  \\
PN    & -424.9502     & 8.2     & 425.9331     & 8.4     & -0.0667     & -72.6    & 6.44$\times 10^{-9}$  & 17  \\
SO$_2$   & -589.7104     & 8.4     & 592.3903      & 8.9     & -0.0636     & -75.0    & -1.00$\times 10^{-12}$     & 127 \\
\hline
Mean      &       & 10.5    &       & 12.4    &       & -75.8    &       & 76  \\
Max      &             &         &             &         &             &          &        & 373 \\
Min      &             &         &             &         &             &          &       & 13  \\
Median      &             &         &             &         &             &          &      & 26 \\
      \hline
\hline
\end{tabular}
\end{table}

\begin{table}[h]
\caption{Energies, kinetic energies, chemical potentials and percentage errors relative to Kohn-Sham LDA exchange-only calculations using the E00 functional. The error in the particle number at convergence and the number of TRIM iterations required to reach convergence are also given.}
\begin{tabular}{lrrrrrrrr}
\hline
\hline
Molecule   & E / $E_\text{h}$          & \%Err E & T / $E_\text{h}$       & \%Err T & $\mu$ / $E_\text{h}$         & \%Err $\mu$ & $N_\text{err}$        & $N_\text{iter}$  \\
\hline
AlF   & -362.4613  & 6.9     & 366.4051   & 8.1     & -0.0489    & -74.7  & -9.54$\times 10^{-10}$ & 28  \\
C$_2$H$_4$  & -84.1801   & 9.5     & 86.3216    & 12.0    & -0.0890    & -57.9  & -1.42$\times 10^{-10}$ & 15  \\
C$_2$H$_4$O & -165.0977  & 9.3     & 168.3587   & 11.3    & -0.0881    & -52.9  & 1.43$\times 10^{-10}$  & 18  \\
C$_3$H$_4$  & -125.0634  & 9.5     & 128.1008   & 12.0    & -0.0902    & -50.8  & 4.39$\times 10^{-10}$  & 16  \\
CH$_2$O  & -121.0809  & 7.5     & 127.8813   & 13.4    & -0.0254    & -86.2  & 7.22$\times 10^{-10}$  & 413 \\
CH$_3$F  & -148.2528  & 7.7     & 154.3348   & 12.0    & -0.0262    & -89.5  & 4.90$\times 10^{-11}$   & 599 \\
CH$_4$   & -43.2835   & 9.5     & 44.6750    & 12.6    & -0.0891    & -70.5  & -1.40$\times 10^{-11}$  & 24  \\
CO    & -119.6121  & 7.2     & 126.4255   & 13.3    & -0.0169    & -94.1  & 1.00$\times 10^{-12}$  & 82  \\
FCCH  & -187.6944  & 8.0     & 194.6610   & 11.9    & -0.0441    & -79.7  & 5.00$\times 10^{-12}$     & 22  \\
FCCH  & -205.0695  & 8.0     & 212.1432   & 11.7    & -0.0373    & -86.7  & 5.00$\times 10^{-12}$     & 24  \\
H$_2$C$_2$O & -161.8091  & 7.9     & 169.7527   & 13.1    & -0.0464    & -73.9  & -5.20$\times 10^{-11}$  & 744 \\
H$_2$O   & -82.0368   & 9.0     & 83.3416    & 10.6    & -0.0838    & -62.2  & 2.50$\times 10^{-11}$   & 35  \\
H$_2$S   & -420.9040  & 6.1     & 421.5687   & 6.2     & -0.0870    & -54.4  & 6.00$\times 10^{-12}$     & 30  \\
HCN   & -100.3770  & 9.4     & 102.1398   & 11.2    & -0.0861    & -70.3  & 2.00$\times 10^{-11}$     & 16  \\
HCP   & -401.6196  & 6.6     & 402.7220   & 6.9     & -0.0867    & -60.8  & 4.00$\times 10^{-12}$     & 23  \\
HFCO  & -229.3959  & 8.9     & 232.2391   & 10.1    & -0.0815    & -66.7  & -8.00$\times 10^{-12}$    & 17  \\
HF    & -107.6172  & 8.5     & 108.4587   & 9.3     & -0.0782    & -74.5  & 4.70$\times 10^{-11}$   & 25  \\
HOF   & -188.6026  & 8.9     & 190.2197   & 9.8     & -0.0753    & -66.1  & -9.00$\times 10^{-12}$    & 25  \\
LiF   & -114.4842 & 8.1     & 116.5371 & 10.1    & -0.0526 & -71.9  & 1.00$\times 10^{-12}$  & 175 \\
LiH   & -8.6248    & 12.0    & 8.6427     & 11.9    & -0.0956    & -25.5  & 2.16$\times 10^{-9}$ & 14  \\
N$_2$    & -117.7391  & 9.3     & 119.7378   & 11.1    & -0.0521    & -84.2  & 2.60$\times 10^{-11}$   & 14  \\
N$_2$O   & -198.5413  & 9.2     & 201.9919   & 11.0    & -0.0603    & -77.4  & 1.30$\times 10^{-11}$   & 18  \\
NH$_3$   & -60.6616   & 9.3     & 62.1522    & 11.7    & -0.0867    & -53.1  & 1.00$\times 10^{-12}$  & 16  \\
O$_3$    & -242.2056  & 9.0     & 247.0222   & 10.6    & -0.0159    & -87.4  & 1.90$\times 10^{-11}$   & 16  \\
OCS   & -538.0475  & 6.1     & 549.0326   & 8.2     & -0.0550    & -76.4  & 1.00$\times 10^{-12}$     & 127 \\
OF$_2$   & -295.0262  & 8.8     & 297.6793   & 9.8     & 0.0069     & -103.0 & 2.00$\times 10^{-12}$     & 16  \\
PN    & -419.0825  & 6.7     & 420.0526   & 6.9     & -0.0818    & -66.3  & 7.00$\times 10^{-12}$     & 27  \\
SO$_2$   & -581.4668  & 6.9     & 584.1112   & 7.4     & -0.0742    & -70.9  & -1.00$\times 10^{-12}$    & 22  \\
\hline
Mean      &    & 8.4     &      & 10.5    &     & -71.0  &       & 93  \\
Max      &            &         &            &         &            &           &        & 744 \\
Min      &            &         &            &         &            &           &      & 14  \\
Median      &            &         &            &         &            &           &     & 24 \\
      \hline
\hline
\end{tabular}
\end{table}